\documentclass[aip,jcp, amsmath,amssymb,
reprint,longbibliography]{revtex4-1}

\usepackage{graphicx}
\usepackage{dcolumn}
\usepackage{bm}

\usepackage[utf8]{inputenc}
\usepackage[T1]{fontenc}
\usepackage{mathptmx}
\usepackage{etoolbox}
\newcommand{\nd}{\noindent}
\usepackage{graphics}
\usepackage[margin=2cm]{geometry}
\usepackage{caption}
\usepackage{subcaption}
\usepackage{color}
\usepackage{multirow}
\usepackage{amsmath}
\usepackage{amsfonts}

\makeatletter
\def\@email#1#2{%
 \endgroup
 \patchcmd{\titleblock@produce}
  {\frontmatter@RRAPformat}
  {\frontmatter@RRAPformat{\produce@RRAP{*#1\href{mailto:#2}{#2}}}\frontmatter@RRAPformat}
  {}{}
}%
\makeatother
\begin{document}

\preprint{AIP/123-QED}

\title[Exploring run-and-tumble movement in confined settings through simulation]{Exploring run-and-tumble movement in confined settings through simulation}
\author{Dario Javier Zamora}
 \affiliation{Dipartimento di Scienza e Alta Tecnologia, Universit{\`a} degli Studi dell’Insubria, Via Valleggio 11, 22100 Como, Italy}
 \affiliation{Instituto de F\'isica del Noroeste Argentino, CONICET and Universidad Nacional de Tucum\'an, Av. Independencia 1800, Tucum\'an, CP 4000, Argentina}
 \email{dariojavier.zamora@uninsubria.it}
\author{Roberto Artuso}
 \affiliation{Dipartimento di Scienza e Alta Tecnologia, Universit{\`a} degli Studi dell’Insubria, Via Valleggio 11, 22100 Como, Italy}
 \affiliation{I.N.F.N Sezione di Milano, Via Celoria 16, 20133 Milano, Italy}

\date{\today}

\begin{abstract}
Motion in bounded domains is a fundamental concept in various fields, including billiard dynamics and random walks on finite lattices, and has important applications in physics, ecology, and biology. An important universal property related to the average return time to the boundary, the Mean Path Length Theorem (MPLT), has been proposed theoretically and experimentally confirmed in various contexts. We investigated a wide range of mechanisms that lead to deviations from this universal behavior, such as boundary effects, reorientation, and memory processes. This study investigates the dynamics of run-and-tumble particles within a confined two-dimensional circular domain. Through a combination of theoretical approaches and numerical simulations, we validate the MPLT under uniform and isotropic particle inflow conditions. The research demonstrates that although the MPLT is generally applicable for different step length distributions, deviations occur for non-uniform angular distributions, non-elastic boundary conditions, or memory processes. These results underline the crucial influence of boundary interactions and angular dynamics on the behavior of particles in confined spaces. Our results provide new insights into the geometry and dynamics of motion in confined spaces and contribute to a better understanding of a broad spectrum of phenomena ranging from the motion of bacteria to neutron transport. This type of analysis is crucial in situations where inhomogeneity occurs, such as multiple real-world scenarios within a limited domain.
\end{abstract}

\maketitle

\section{\label{sec:level1}Introduction}
The properties of motion within bounded domains are applicable across a variety of fields such as physics, ecology, and biology. Motion in bounded domains has far-reaching implications, from fundamental mathematical insight to practical applications in environmental science, biology, medicine, and ecology. This lays the foundation for predicting and controlling the behavior of particles or agents in diverse and restricted environments, offering potential benefits to numerous scientific and technological fields.

A key aspect of motion in confined environments is the Mean Path Length Theorem (MPLT), a generalization of a geometric result originally proposed by Cauchy \cite{Cauchy1850}. The (MPLT) is a fundamental result in the study of particle dynamics in confined domains, providing a universal expression for the average path length of particles. In particular, this theorem states that the mean path length $<L>$ of particles within a bounded domain $\Omega$ is determined by the ratio between the volume $V_\Omega$ and the surface area $\Sigma_\Omega$, regardless of the details of the dynamics of the particles or the shape of the domain.

\begin{equation}
    <L> = \eta_d \frac{V_\Omega}{\Sigma_\Omega},
    \label{Cauchy}
\end{equation}

\nd where $\eta_d$ is a numerical factor that depends only on the space dimensionality $d$ (in this case, we consider in detail $d=2$ and $\eta_2=\pi$).

More precisely, the MPLT concerns the case in which $<L>$ is the mean length of randomly distributed chords intersecting a convex body. Based on the linear Boltzmann equation, the MPLT was also extended to scenarios in which the particles do not move in straight lines but are scattered \cite{Bardsley1981}. This indicates that the Cauchy formula applies more generally to the stochastic paths of Pearson random walks with exponentially distributed flight lengths. This demonstrates that under the conditions of a uniform and isotropic particle influx, the absence of absorption or particle reproduction within the medium, and a detailed balance in scattering probabilities, the average path length remains unchanged. The analysis contributes to the understanding of transport phenomena in scattering media, which are relevant, for example, for the estimation of reaction rates in neutron transport \cite{Kruijf2003,Dirac1943,Case1953}, ionic recombination in gases and other processes in finite volumes. Furthermore, this theorem has recently been generalized for non-convex domains and extended to the case of Brownian motion \cite{Blanco2003, Mazzolo2004, Santalo1976}. The work of Benichou et al. \cite{Benichou2005} discusses the averaged residence times of stochastic motions within bounded domains, focusing on the mean first exit and residence times for particles undergoing random motions, such as Pearson random walks, by employing a backwards Chapman Kolmogorov equation. This study is motivated by various physical systems, including animal trajectories \cite{Jeanson2003} and neutron scattering processes.  The case of arbitrary closed trajectories in arbitrary domains was investigated by Caballero et al. \cite{Caballero2022}, and it is shown that the mean path length still agrees with the Cauchy formula as long as no closed trajectory is completely contained in the domain. Shukla et al. \cite{Shukla2019} extend the Cauchy formula to lattice-based random walks, providing a straightforward understanding compared to previous models relying on continuum-based Boltzmann transport equations. By analyzing random walks on an $N \times N$ square lattice, the research provides both numerical results and theoretical underpinnings of the observed phenomena, suggesting that the Cauchy formula can be generalized beyond traditional applications to include more complex random walk scenarios. This property holds for anomalous and classical photon propagation \cite{Binzoni2022}. All these generalizations imply that the Cauchy formula, Eq. (\ref{Cauchy}), is applicable to a wide range of scenarios and phenomena, hence the importance. Cauchy's universality was recently confirmed by experimental results on the movement of bacteria \cite{Frangipane2019} and the propagation of light in scattering media \cite{Pierrat2014,Savo2017}.

In addition, knowing the statistics of random walks in confined settings offers insights into a range of phenomena, such as the conditions under which active particles can escape barriers or be trapped by external forces, or how chemotactic fields influence particle motion towards or away from high nutrient concentrations, applicable to bacterial motion, intracellular transport, and animal foraging, bridging the gap between theoretical models and real-world biological processes \cite{Angelani2014, Rashid2018, Long2017}. These systems present what is called a Run-and-tumble movement, characterized by randomly reoriented ballistic movements. Understanding the mean chord length can provide foundational insights into the geometry and scale of motion within a domain, which is indirectly relevant to estimating escape times or paths in narrow escape scenarios, especially when considering the impact of domain geometry on escape dynamics \cite{Paoluzzi2020, Ruppretch2015, Singer2006} or target searching \cite{Ruppretch2016}.

It is widely recognized that numerous stochastic transport phenomena observed in the real world do not exhibit jump lengths that are exponentially distributed, nor do they show scattering angles that are uniformly random \cite{Davis2004,Barthelemy2008,Svensson2014}. For instance, isotropic random flights having fixed length have applications in polymer chains \cite{Flory1969}, and an exponential law has applications, as mentioned, to neutron diffusion processes and locomotion problems in biology. It naturally raises the question of whether these processes still adhere to similar universal characteristics. In fact, heuristic reasoning indicates that Cauchy's formula is applicable to any (isotropic) random walk as long as the random walk's entry into the domain involves a length distribution that aligns with equilibrium \cite{Benichou2005, Blanco2006, Mazzolo2014}. In particular, some attention was recently given in the literature to the role of boundary conditions in influencing the absorption time of run-and-tumble particles \cite{Angelani2015, Artuso2024}, and to the particle's speed distribution or the spatial dimensionality \cite{Mori2020}.

Complex dynamics in bounded domains are occasionally linked to non-homogeneous distributions, as in the case of active particles \cite{Caprini2019,Volpe2014,Yang2014} or when interactions with walls are present \cite{Souzy2022}. Therefore, it is natural to wonder if Cauchy universality is upheld in these situations. This is the main topic we will cover, giving examples of mechanisms causing nonuniform stationary probability distributions and breaking Cauchy universality; we will take into account both the scenario where particles move in a stochastic way inside the body (random walk) and the scenario where they follow rectilinear trajectories between consecutive collisions with the boundary (billiard). Through theoretical analysis and numerical simulations, this study examines how different boundary and reorientation conditions affect the motion's universal properties. This research offers potential implications for understanding motion in confined systems across various scientific domains, extending beyond the classical applications of billiards and random walks to include biological and physical phenomena where such conditions are prevalent.

The paper is structured as follows. In the next section, the methodology for simulating particle movement within a 2D circular space is described, including variations in step length and angular distributions under different boundary conditions. Section \ref{validation} presents the analysis confirming the MPLT under uniformly random angular distributions and highlights deviations with non-uniform distributions. The section \ref{distributions:sec} examines path length distributions and their variance, illustrating the transition from diffusive to quasi-ballistic movements. Section \ref{radial:sec} explores the radial probability distributions, showing uniformity under elastic boundary conditions and deviations with random angle conditions. In section \ref{nemo:sec}, we introduce memory effects in particle reorientation and their impact on path length and radial distributions. Finally, we summarize the key findings and their implications in section \ref{conclusions}.

\section{Simulations}
\label{simulations}

\begin{figure*}[!]
    \centering
    \begin{subfigure}[b]{0.33\textwidth}
        \centering
        \includegraphics[width=\linewidth]{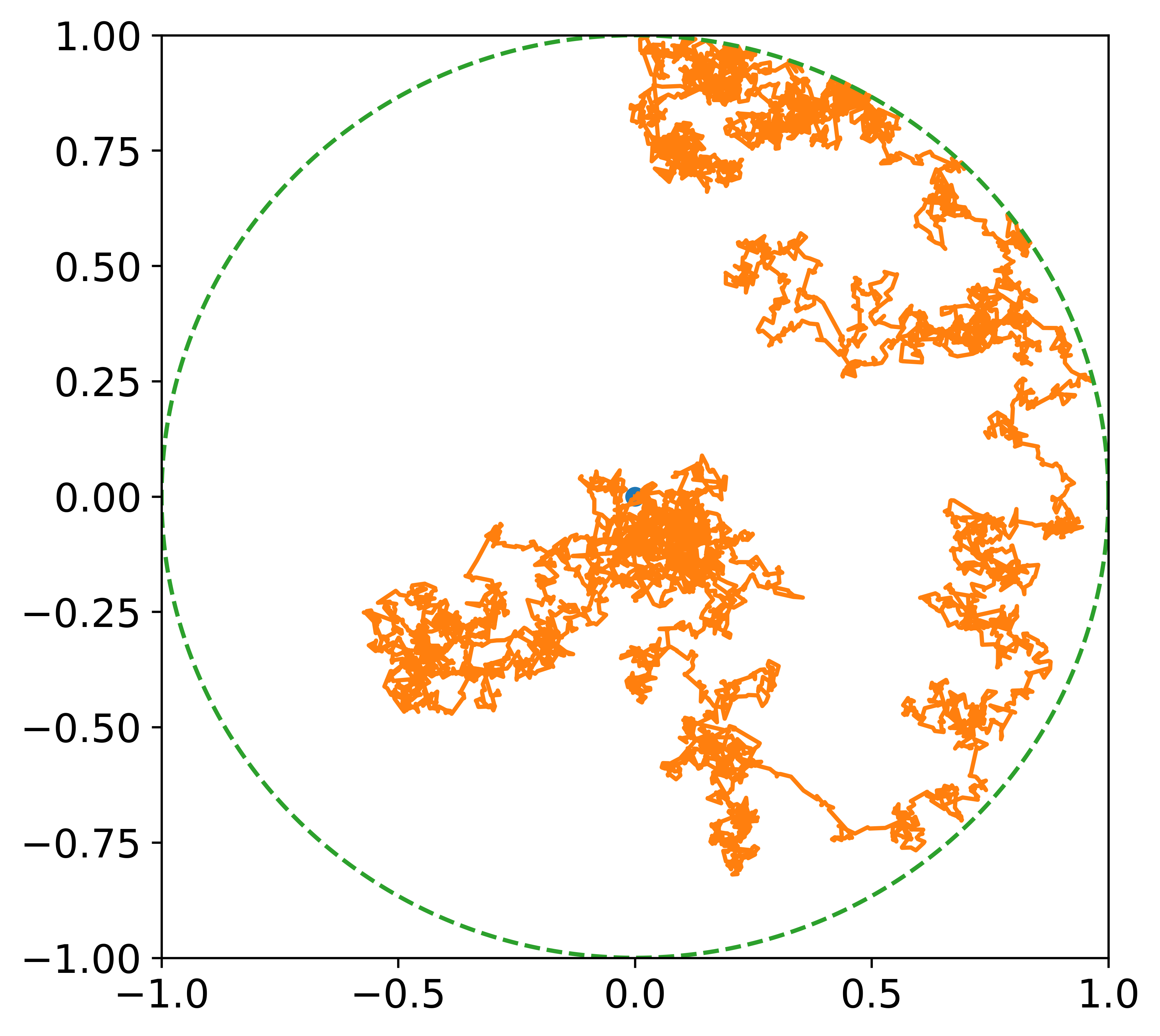}
        \caption{}
    \end{subfigure}
    \begin{subfigure}[b]{0.33\textwidth}
        \centering
        \includegraphics[width=\linewidth]{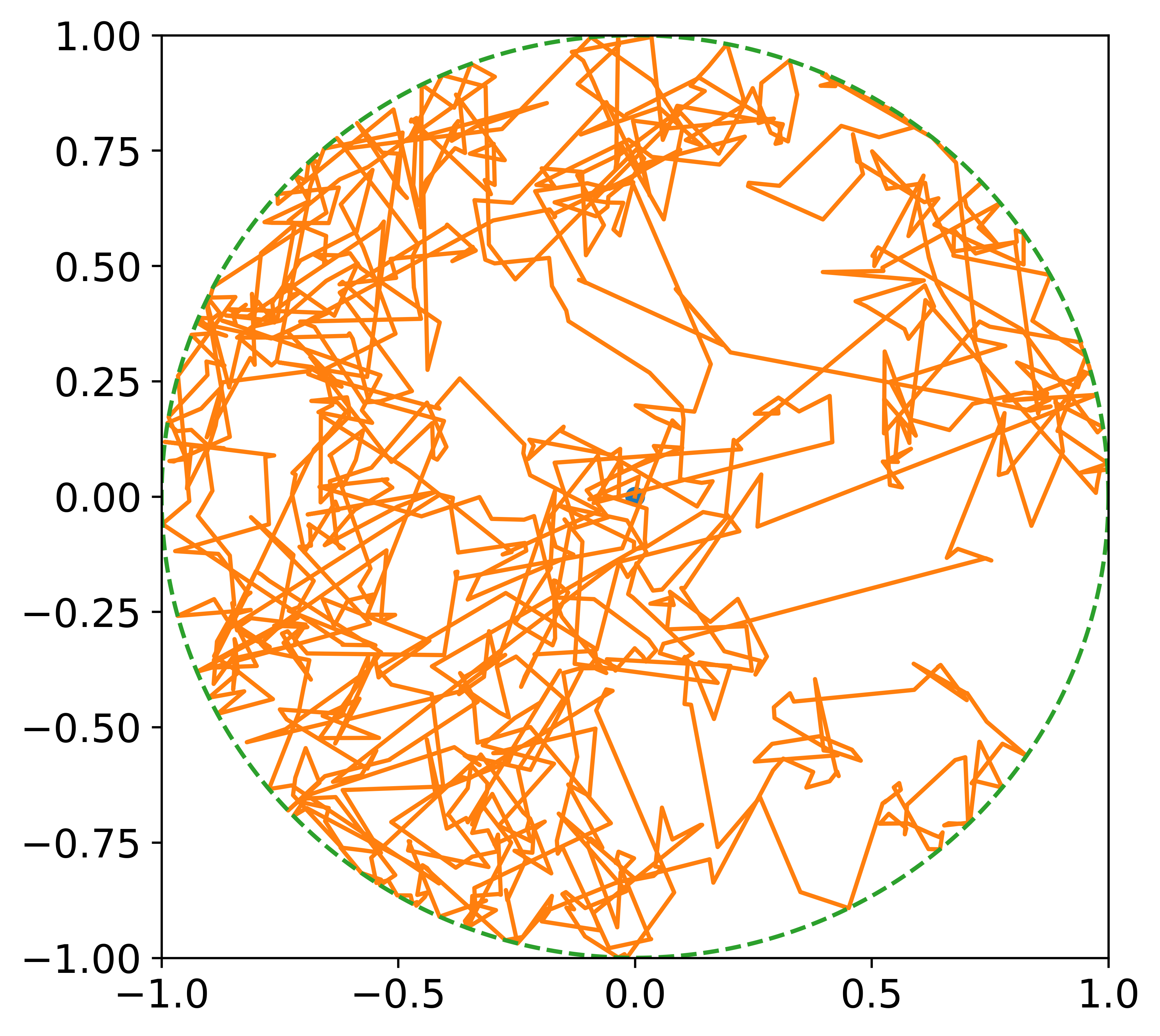}
        \caption{}
    \end{subfigure}
    \\
    \begin{subfigure}[b]{0.33\textwidth}
        \centering
        \includegraphics[width=\linewidth]{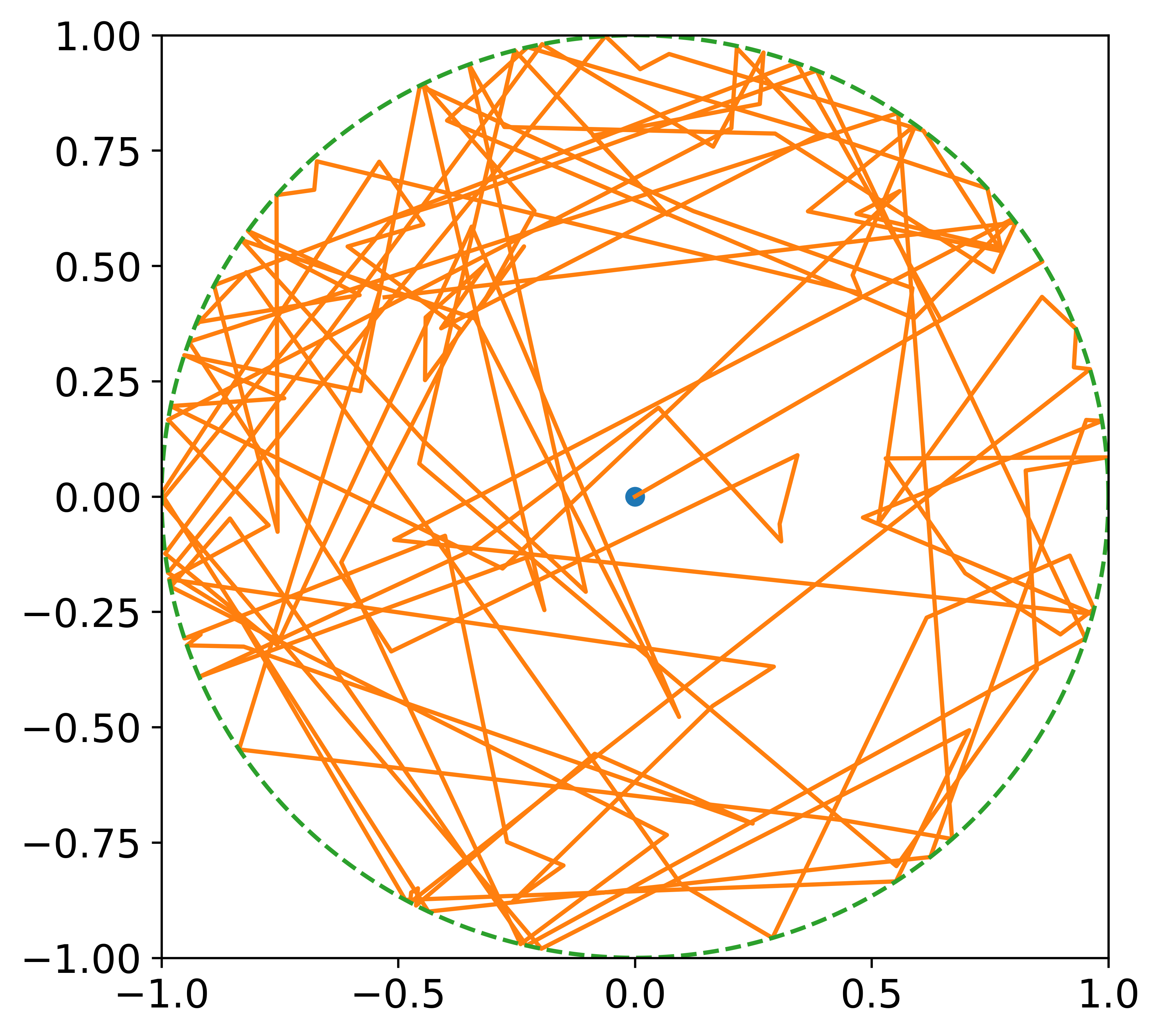}
        \caption{}
    \end{subfigure}
    \begin{subfigure}[b]{0.33\textwidth}
        \centering
        \includegraphics[width=\linewidth]{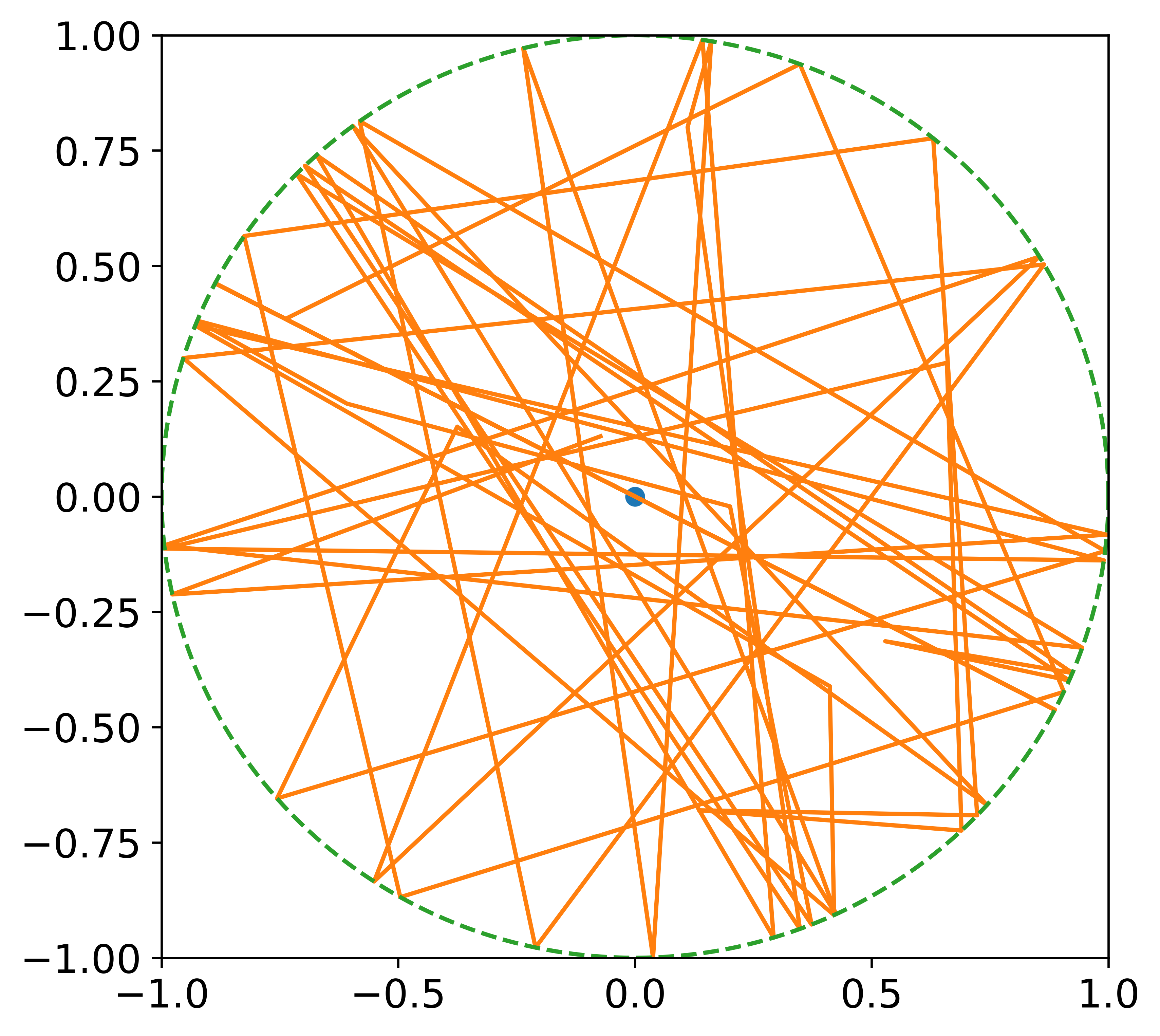}
        \caption{}
    \end{subfigure}
    \caption{Trajectory of the particle inside a circle with elastic boundary. $l/R$: (a) $l/R=0.01$, (b) $l/R=0.1$, (c) $l/R=1$, (d) $l/R=10$. Figure (a) is diffusive, on the other extreme, (d) is quasi-ballistic.}
    \label{trajectories}
\end{figure*}

To replicate a simple geometric setting common to experimental realizations \cite{Rashid2018, Long2017, Frangipane2019, Souzy2022}, our code simulates a particle moving within a unit 2D circular space. In our model, the particle travels at a constant unit speed, $v=1$. Therefore, the time between successive collisions is equal to the total trajectory length, allowing us to discuss average residence time and mean path length interchangeably. In two dimensions, the mean path length, as defined in eq. (\ref{Cauchy}), is expressed as follows:

\begin{equation}
	<L>=\pi\frac{S}{P},
\end{equation}

\nd where S is the surface and P the perimeter of the domain. So, for the disk of radius $R=1$ we have 

\begin{equation}
    <L> = \frac{\pi}{2} R = \frac{\pi}{2}.
\end{equation}

After each step (run), the particle changes direction (tumble). Consequently, to define the Run-and-Tumble dynamics, we must specify: (i) the distribution of step lengths, $Q(\ell)$; (ii) the angular distribution of the tumble events; and (iii) the outgoing angle after a collision with the boundary. In this work, we explore the statistics of the run-and-tumble movement under four conditions of step length distribution, namely: (a) Gaussian distributed (as in some standard random walks),

\begin{equation}
	Q(\ell) = \frac{1}{\sigma\sqrt{2\pi}}\exp\left(-\frac{1}{2}(\frac{\ell-l}{\sigma})^2\right),
\end{equation}

\nd where $l=<\ell>$ is the mean step length and $\sigma$ is the standard deviation, (b) uniformly random, (c) fixed (as in polymer chains \cite{Flory1969}), and (d) exponential distributed (typical of bacteria mobility \cite{Frangipane2019} and neutron scattering \cite{Kruijf2003}),

\begin{equation}
	Q(\ell) = \frac{1}{l}\exp\left(-\frac{\ell}{l}\right).
\end{equation}

In the case of the Gaussian distribution, no cutoff was used. This means that there is a possibility that the particles move backwards if $\ell$ takes a negative value. We will analyze the statistics and the validation of the MPLT under various angular distributions: (a) a fixed angle, which can be considered as a representation of strong chemotaxis (preferred direction); (b) a Gaussian distribution, representing weak chemotaxis, since there is a preferred direction with some deviation, a similar approach is used in the work of Romanczuk and Salbreux \cite{Romanczuk2015}; (c) the particle is allowed to move in any direction uniformly distributed in $[0, 2\pi]$, which is the typical case considered in most studies.

When the particle reaches the boundary, it changes its direction of motion as a result of the collision, and the new direction may depend on the angle of incidence. In this paper, we will consider two different boundary conditions: 1- elastic (i.e., the angle of reflection is equal to the angle of incidence), 2- the particle exits the collision with the boundary at a random outgoing angle.

In our simulation with elastic boundaries, when a particle takes a step that would end outside the circular boundary, we detect this by calculating the distance between the new position and the center of the circle. If this distance exceeds the radius of the circle, we consider the particle to have collided with the boundary. Upon detecting such a collision, we compute the point of intersection between the particle's trajectory and the circle boundary. The particle's trajectory is then reflected at this boundary point. The reflection is implemented by calculating the new direction of the particle after the collision, following the principle that the angle of incidence equals the angle of reflection. The particle is then moved in this new direction, ensuring it stays within the boundary. The distance traveled by the particle after the collision is calculated as the difference between the original step length and the distance already traveled between the particle's previous position and the boundary. If this new step also leads the particle outside the boundary, the process is repeated until the particle's position lies within the circle after the reflection. In the case of random outgoing angle, after determining the collision point, the particle is assigned a new random direction after the collision. This is done by drawing a new angle uniformly from the range $[0,2\pi]$.

By adjusting the mean step length, $l$, we can observe different types of movement, as illustrated in Fig. (\ref{trajectories}). While these examples start from the center of the circle, the findings presented in this work are not dependent on the starting location, including points on the boundary or any other random initial positions within the circle. During simulations with extremely few steps, different initial distributions could potentially influence the results, although we verified that the long-term behavior, particularly in the case of many collisions, is less sensitive to the initial distribution of walkers. In Fig. (\ref{trajectories}), the exit angle after each tumble event is uniformly random, and the step length follows an exponential distribution. We observe various movement types according to the ratio $l/R$, where $R$ is the circle's radius. When $l/R<<1$, the movement is diffusive, but when $l/R>>1$, the movement becomes quasi-ballistic. Purely ballistic movement occurs in the limit as $l/R\to\infty$. In our simulations, a combination of both types of movements is always present. We refer to these limit movements as diffusive and quasi-ballistic based on the qualitative behavior of the system: for small $l/R$ values, the predominant movement resembles a random walk, whereas when $l/R$ is large, the predominant movement resembles a billiard (where the particle bounces inside the circle following straight paths).

\section{Mean path length theorem}
\label{validation}

\begin{figure}[h!]
    \centering
    \includegraphics[width=\linewidth]{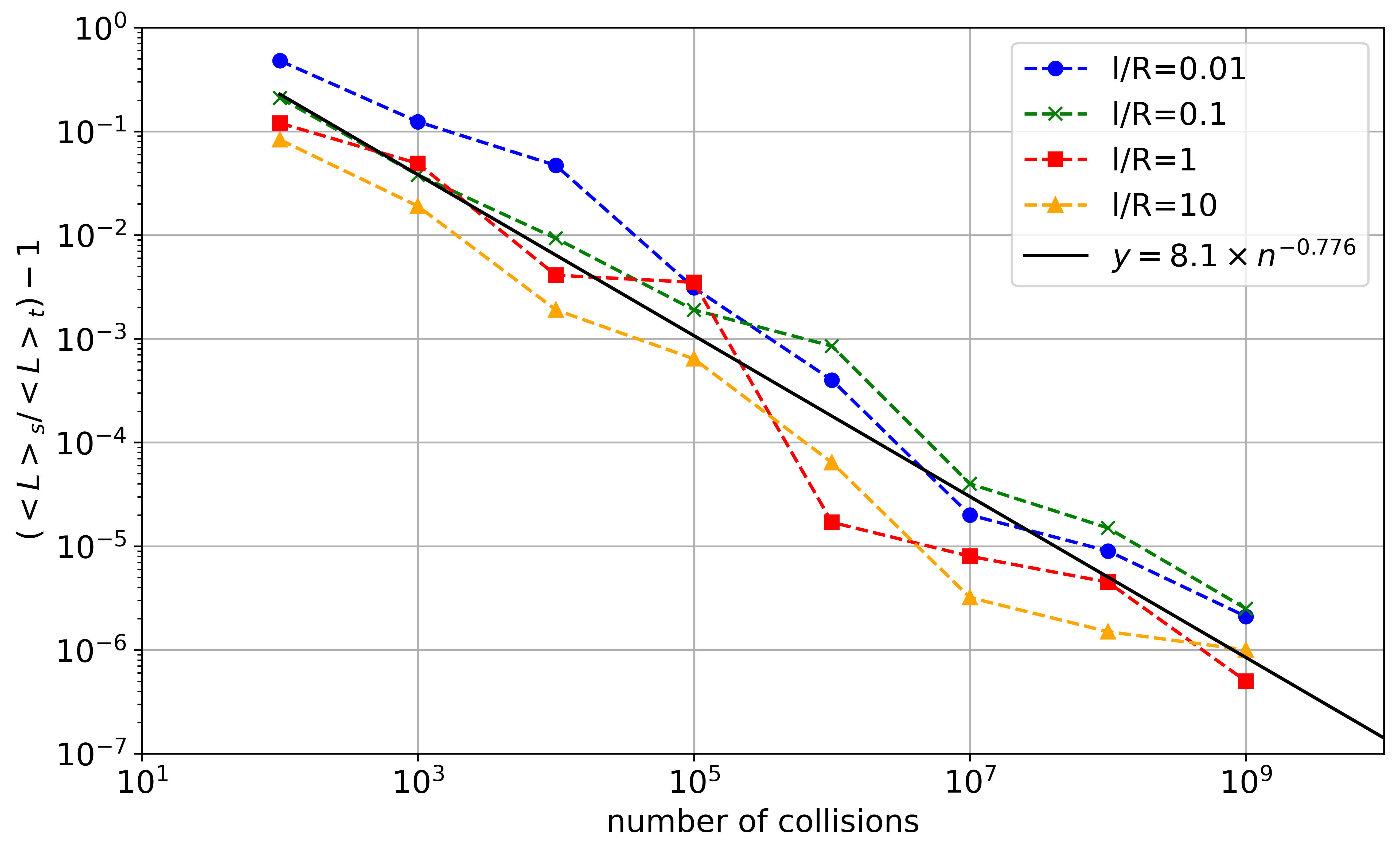}
    \caption{Difference between mean path length according to the theorem and the simulation versus the number of collisions, n, with elastic boundary condition, exponential step length distribution, and uniformly random angular distribution. Each point is an average of 20 simulations. The black solid line is a power-law fitting of the points.}
    \label{meanpathlength}
\end{figure}

First, we explore the conditions under which the Mean Path Length Theorem (MPLT) is valid. To do this, we calculate the mean path length using various combinations of step length and angular reorientation distribution. This analysis is conducted with elastic boundary conditions for four distinct values of $l/R$: $0.01$, $0.1$, $1$, and $10$. We find that the MPLT is valid only when the angular distribution is uniformly random, across all step length distributions. At first, the fact that the mean path length is the same for different values of $l/R$ seems very surprising, because if we replace straight segments (or arcs) with irregular paths, we intuitively expect a twofold mechanism to profoundly change the dynamics (stochastic short returns to the boundary and long walks in the domain without touching the boundary) \cite{Blanco2006}. This fact is confirmed by plotting the ratio $(<L>_s/<L>_t)-1$, where $<L>_s$ represents the mean path length from the simulation, and $<L>_t$ is the theoretical value predicted by the MPLT. If the MPLT is accurate, the simulated and theoretical mean path lengths will converge as the number of collisions increases. Here, we present some of the most illustrative cases.

\begin{figure}[h!]
    \centering
    \begin{subfigure}[b]{0.49\textwidth}
        \centering
        \includegraphics[width=\linewidth]{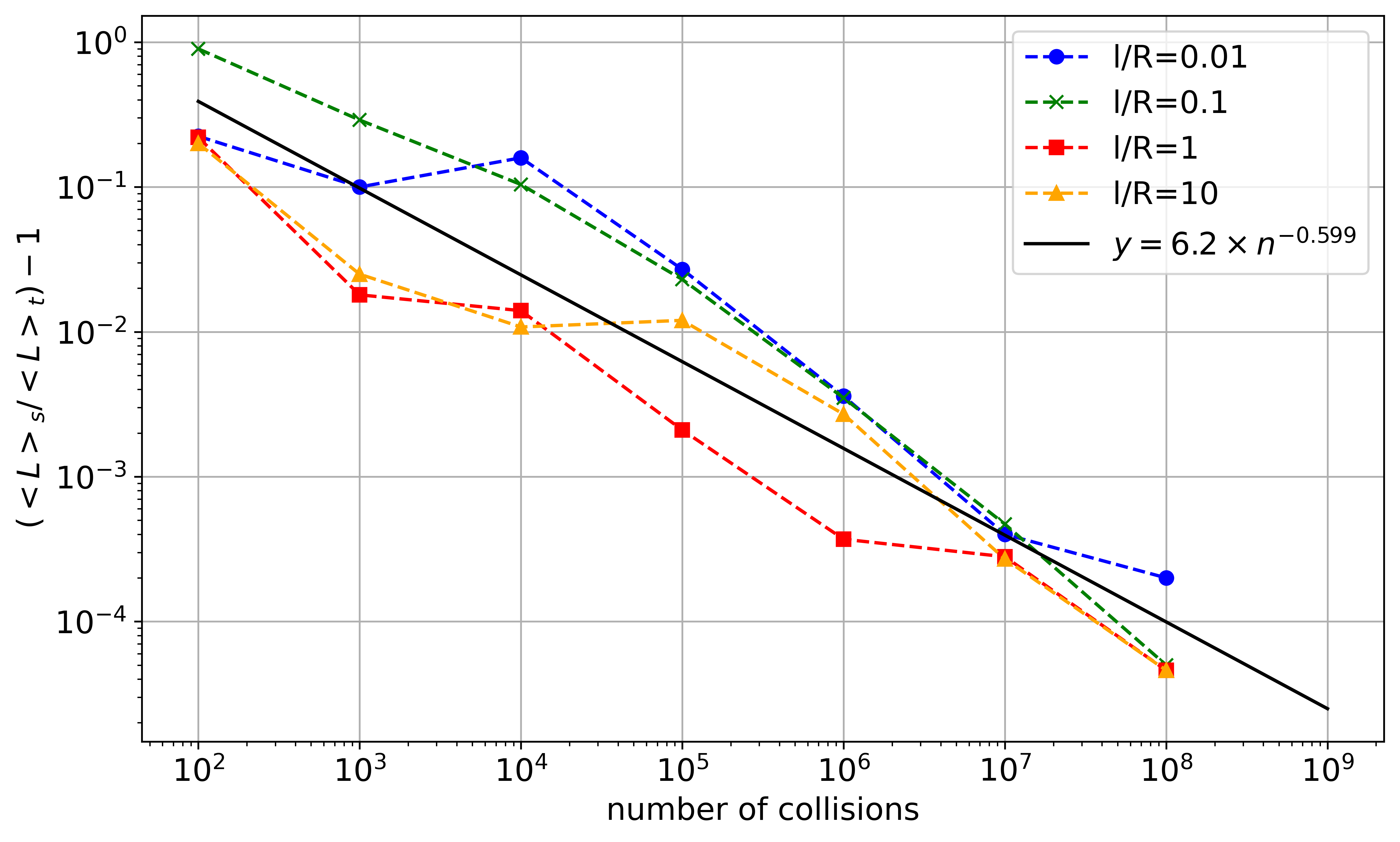}
        \caption{}
        \label{meanpathlengthstepuniform}
    \end{subfigure}
    \hfill
    \begin{subfigure}[b]{0.49\textwidth}
        \centering
        \includegraphics[width=\linewidth]{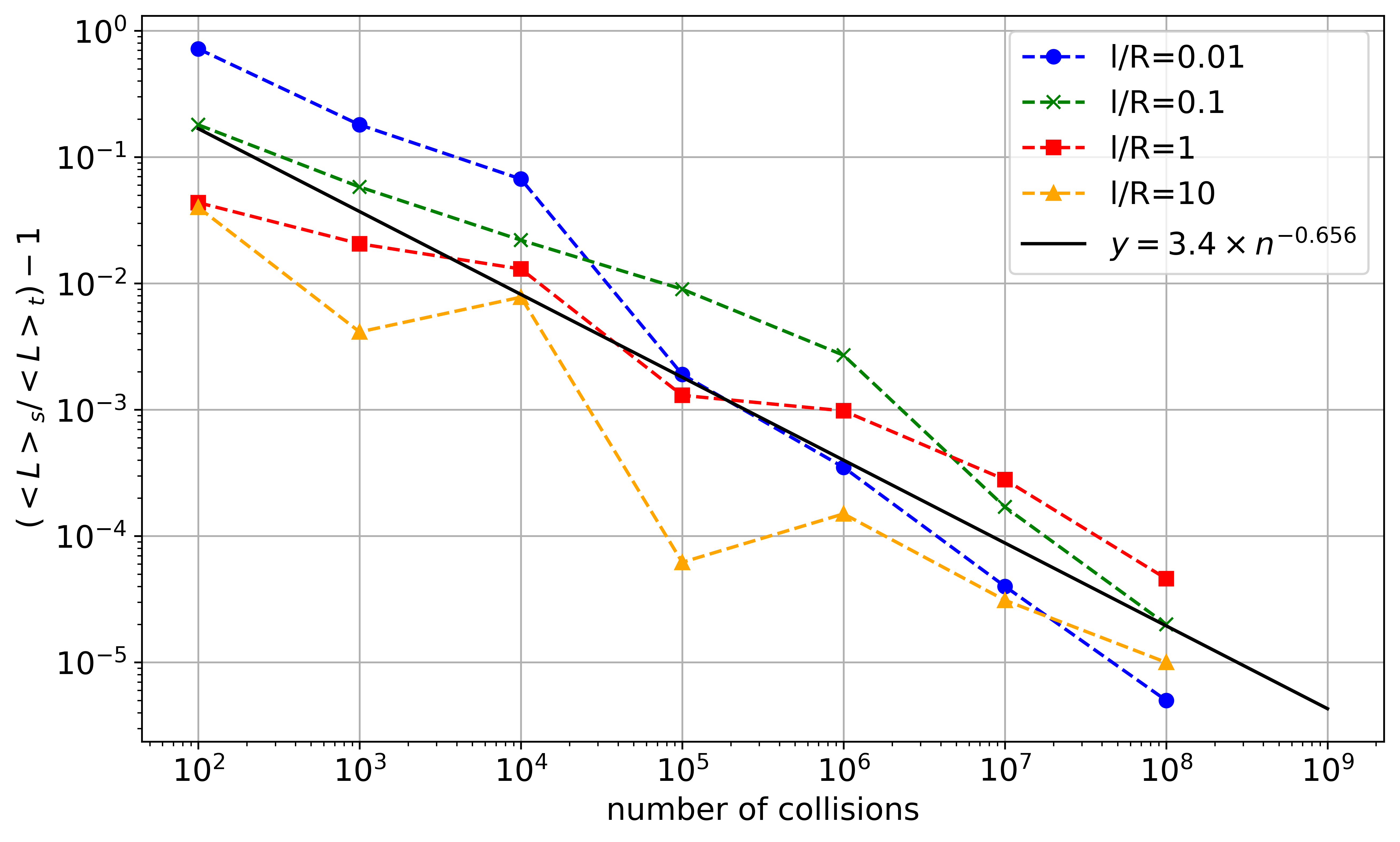}
        \caption{}
        \label{meanpathlengthstepfixed}
    \end{subfigure}
    \caption{Difference between theoretical mean path length and simulation versus the number of collisions with elastic conditions at the boundary, random angle distribution, and (a) random step length between 0 and $2l$, (b) constant step length $l$. Black lines: power-law fittings.}
\end{figure}

In Fig. (\ref{meanpathlength}), a typical graph of $(<L>_{s}/<L>_{t})-1$ is shown for various $l/R$ ratios, using an exponential step length distribution and a uniformly random angular distribution. It is observed that the difference between simulated and theoretical values tends to zero as the number of collisions increases. The step length distribution can be adjusted to any of the mentioned scenarios, and the MPLT will still hold true. For illustration, in Figs. (\ref{meanpathlengthstepuniform}) and (\ref{meanpathlengthstepfixed}), we present two distinct cases of step length distribution: (a) random between $0$ and $2l$, and (b) a constant value of $l$. The differences between the simulated and theoretical values of mean path length approach zero as the number of collisions tends towards infinity in all instances.

As illustrated in the figures \ref{meanpathlength}, \ref{meanpathlengthstepuniform} and \ref{meanpathlengthstepfixed}, the quantity $(<L>_{s}/<L>_{t})-1)$ decreases following a power-law as the number of collisions increases. The black line in the figures are power-law fittings. According to the Central Limit Theorem, under appropriate conditions, the distribution of the sample mean converges to a normal distribution \cite{Montgomery2014}. This holds even if the original variables themselves are not normally distributed. We expect that the speed of convergence is at least of the order $1/\sqrt{N}$: indeed in our case this is not a rigorous statement, since we did not study in detail eventual correlations between successive steps.

\begin{figure}[h!]
    \centering
    \begin{subfigure}[b]{0.49\textwidth}
        \centering
        \includegraphics[width=\linewidth]{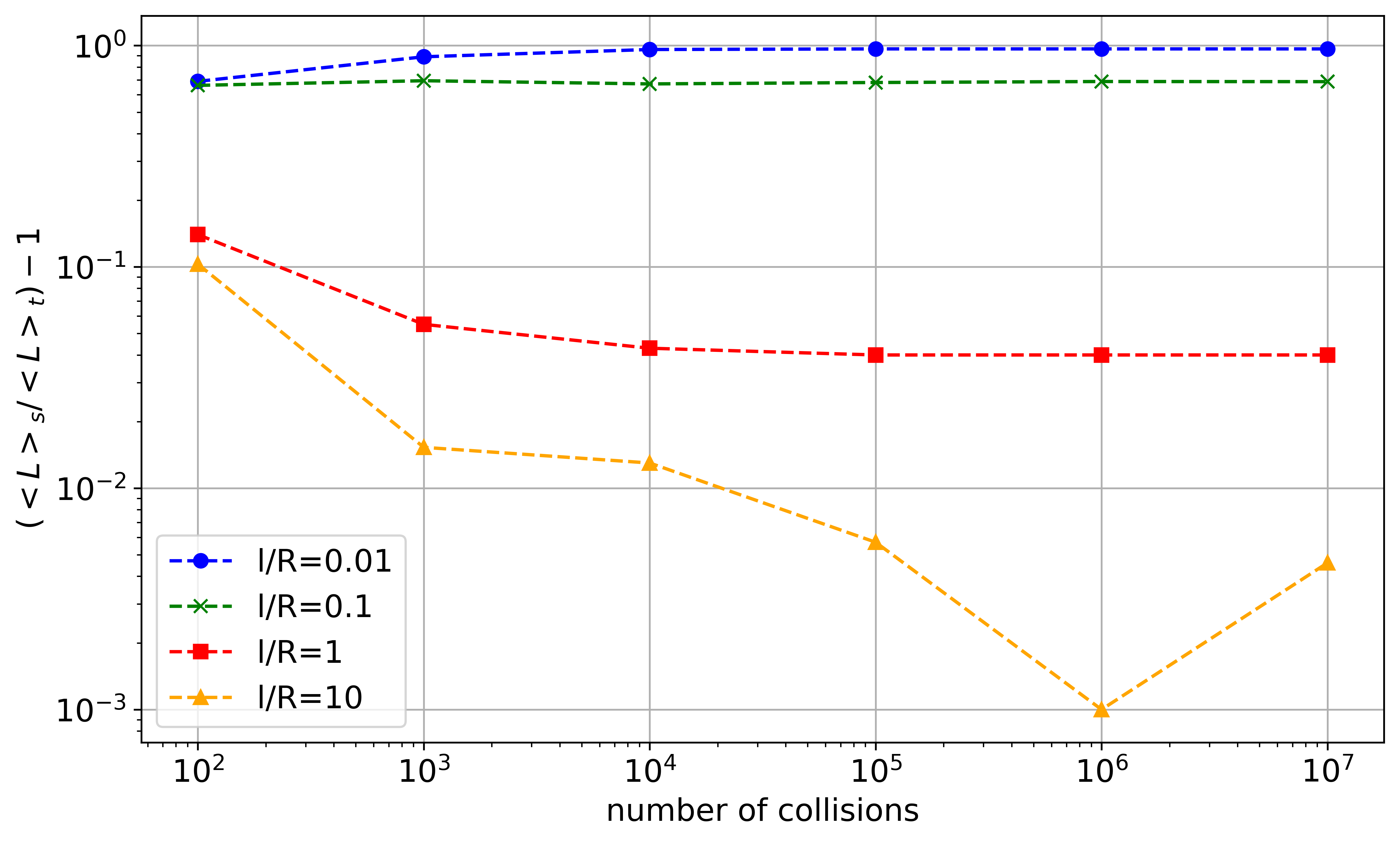}
        \caption{}
        \label{meanpathlengthanglegauss}
    \end{subfigure}
    \hfill
    \begin{subfigure}[b]{0.49\textwidth}
        \centering
        \includegraphics[width=\linewidth]{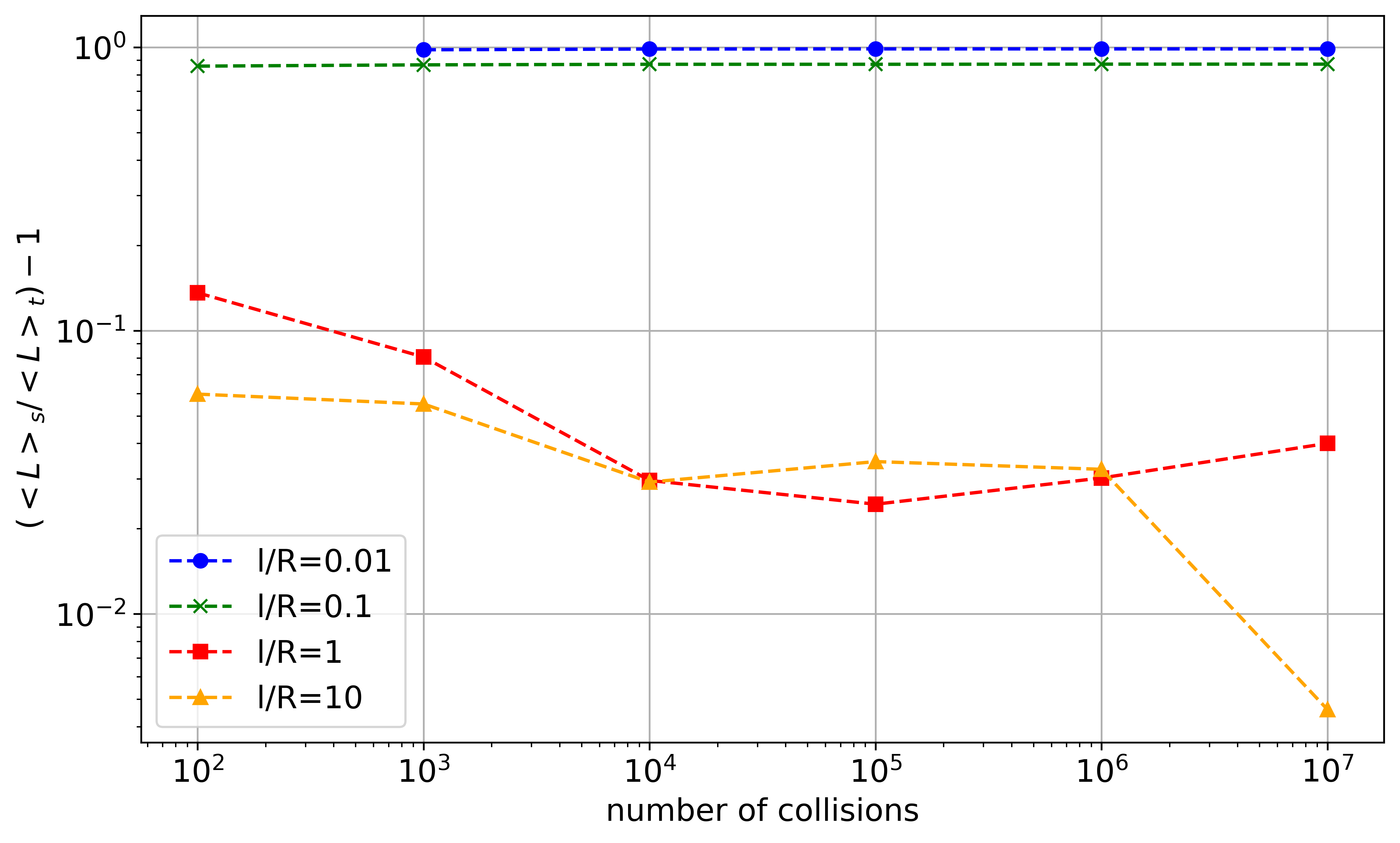}
        \caption{}
        \label{meanpathlengthanglefixed}
    \end{subfigure}
    \caption{Difference between theoretical mean path length simulation versus the number of collisions with elastic conditions at the boundary, exponential step length distribution, and (a) gaussian angle distribution with mean 0, and $\sigma=\pi/2$; (b) fixed value of angle at $\pi/4$.}
\end{figure}

\begin{figure}[h!]
    \centering
    \begin{subfigure}[b]{0.33\textwidth}
        \centering
        \includegraphics[width=\linewidth]{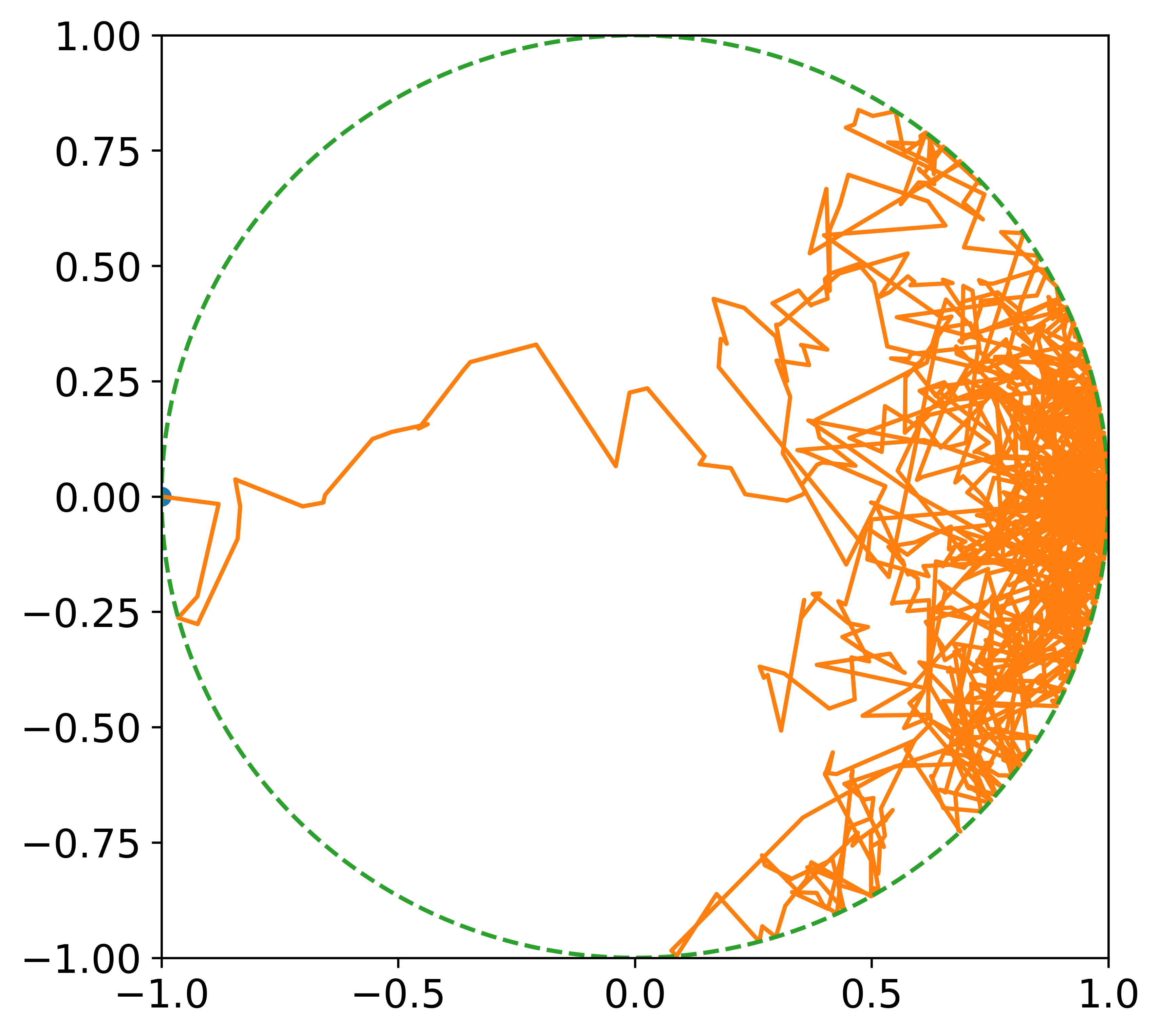}
        \caption{}
    \end{subfigure}
    \begin{subfigure}[b]{0.33\textwidth}
        \centering
        \includegraphics[width=\linewidth]{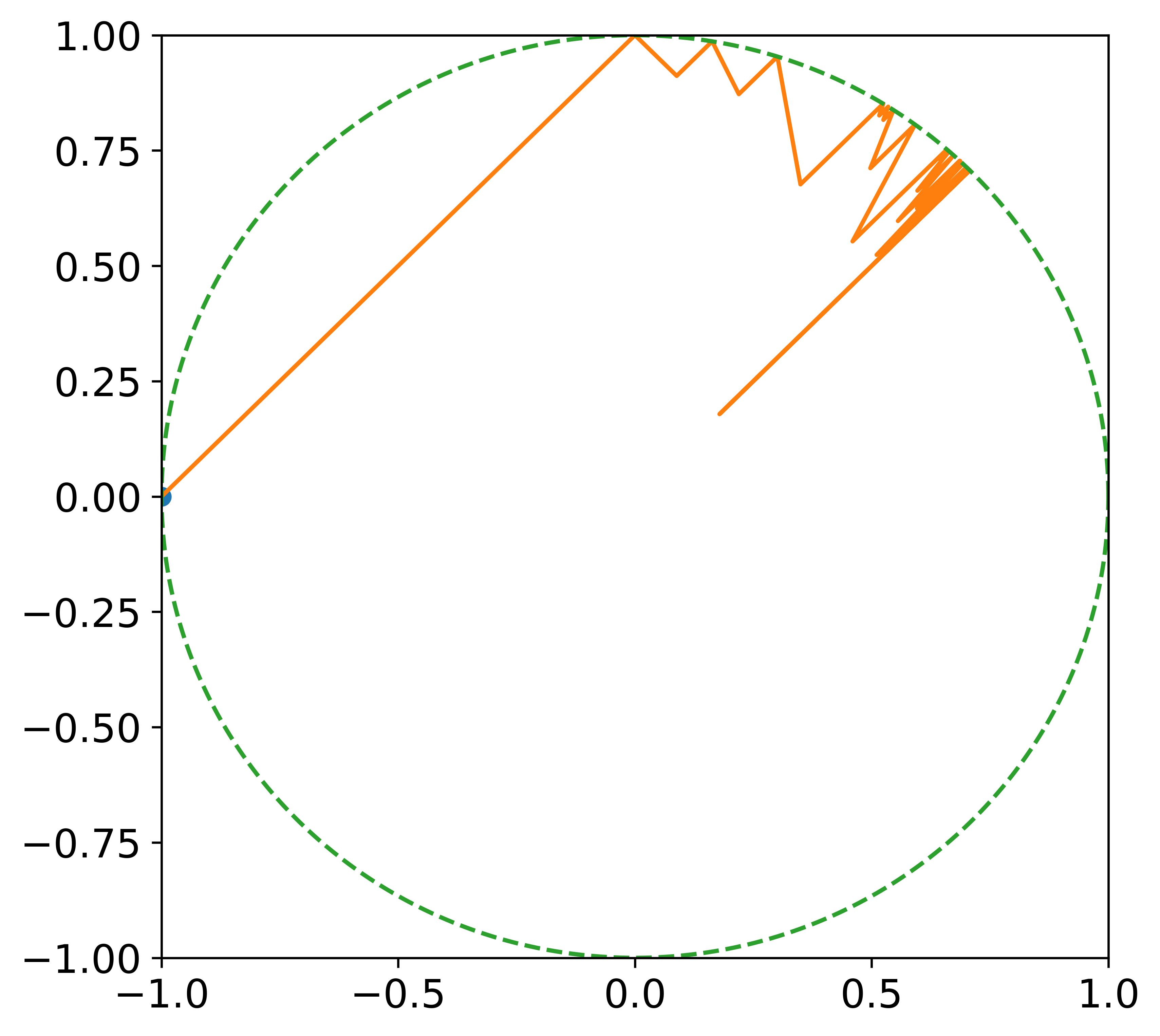}
        \caption{}
    \end{subfigure}
    \caption{Trajectory of 1000 steps of the particle inside a circle with elastic boundary, exponential step length distribution, and $l/R=0.1$. (a) The angle distribution normal-distributed around $\theta=0$, (b) Angle distribution fixed at $\pi/4$.}
    \label{anisotropy}
\end{figure}

On the other hand, if we alter the distribution of the reorientation angle after the tumble events, the assumption of homogeneous flux at the boundary no longer applies, and consequently, the mean path length does not converge to the same value for all $l/R$ ratios, as illustrated in Figs. (\ref{meanpathlengthanglegauss}) and (\ref{meanpathlengthanglefixed}). In Fig. (\ref{meanpathlengthanglegauss}), the angles after the tumble event follow a Gaussian distribution with a mean of $0$ and a standard deviation of $\pi/2$. We observed similar outcomes with various combinations of means and standard deviations. Similarly, Fig. (\ref{meanpathlengthanglefixed}) displays results for a fixed angle of $\pi/4$, yet comparable results are achieved with different preferred direction values.
The figures suggest that the difference between the simulated and theoretical mean path length is greater when the parameter $l/R$ is small, i.e., in the diffusive regime. A diffusive regime is characterized by many small, local changes in position. Initially, the drift component acts to "straighten out" the overall trajectory, but when the boundary is reached, it causes the particle to remain in a specific region of the domain, as seen in Fig. \ref{anisotropy}, introducing significant anisotropy. 
In contrast, in a ballistic regime, particles move in nearly straight lines over longer distances between collisions, with little deviation in their trajectory. When a drift component is introduced into a ballistic system, the directional bias in the movement is less influential, because, from time to time, the particle takes a step long enough to escape from the preferred region of the domain, reducing the anisotropy. This leads to a smaller deviation from the MPLT predictions. The MPLT assumes a certain level of isotropy in the way particles sample the space, and as a result, the average path length measured from simulations deviates more significantly in the diffusive+drift case than in the ballistic+drift case.

The fixed angle and the Gaussian distribution around an angle create a preferred direction, observable in some experimental behaviors of bacteria due to mechanisms like chemotaxis. The fixed angle constitutes a strong condition (the particle consistently moves in a specific direction), while the Gaussian distribution represents a less stringent condition (there is a preferred direction, yet a certain probability exists for the particle to move in another direction close to the preferred one). In scenarios with fixed and Gaussian-distributed angles, the particle tends to collide with the boundary consistently in the same region, as illustrated in Fig. (\ref{anisotropy}). This results in non-uniformity and anisotropy, thus violating the assumptions underlying the MPLT. This outcome is not surprising, as previous works have already clarified the conditions of the angular distribution under which the theorem holds \cite{Bardsley1981, Mazzolo2004, Benichou2005, Mazzolo2014}.

\begin{figure}[h!]
    \centering
    \begin{subfigure}[b]{0.49\textwidth}
    	\centering
    	\includegraphics[width=\linewidth]{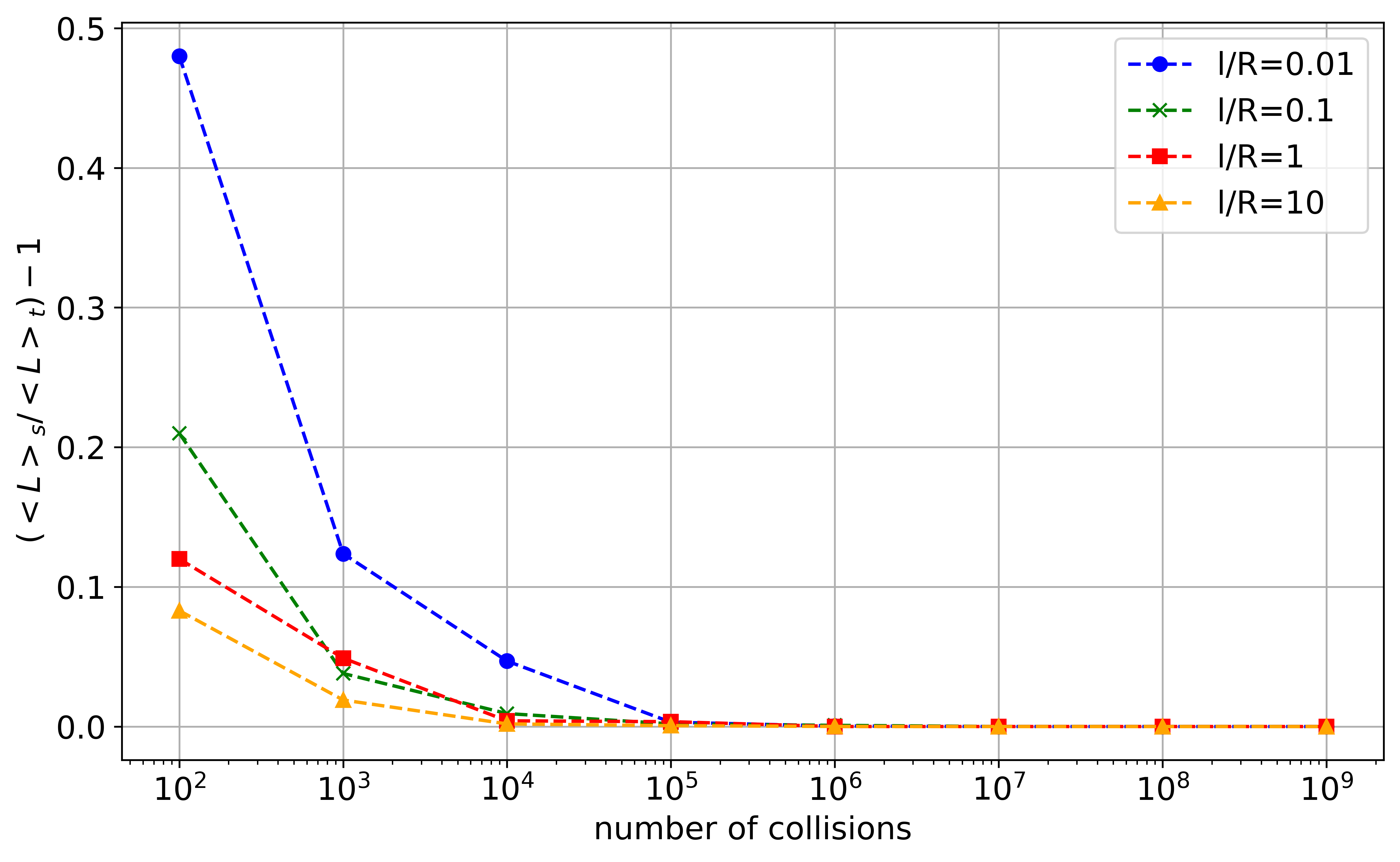}
	\caption{}
	\label{meanpathlengthreflective}
    \end{subfigure}
    \hfill
    \begin{subfigure}[b]{0.49\textwidth}
        \centering
        \includegraphics[width=\linewidth]{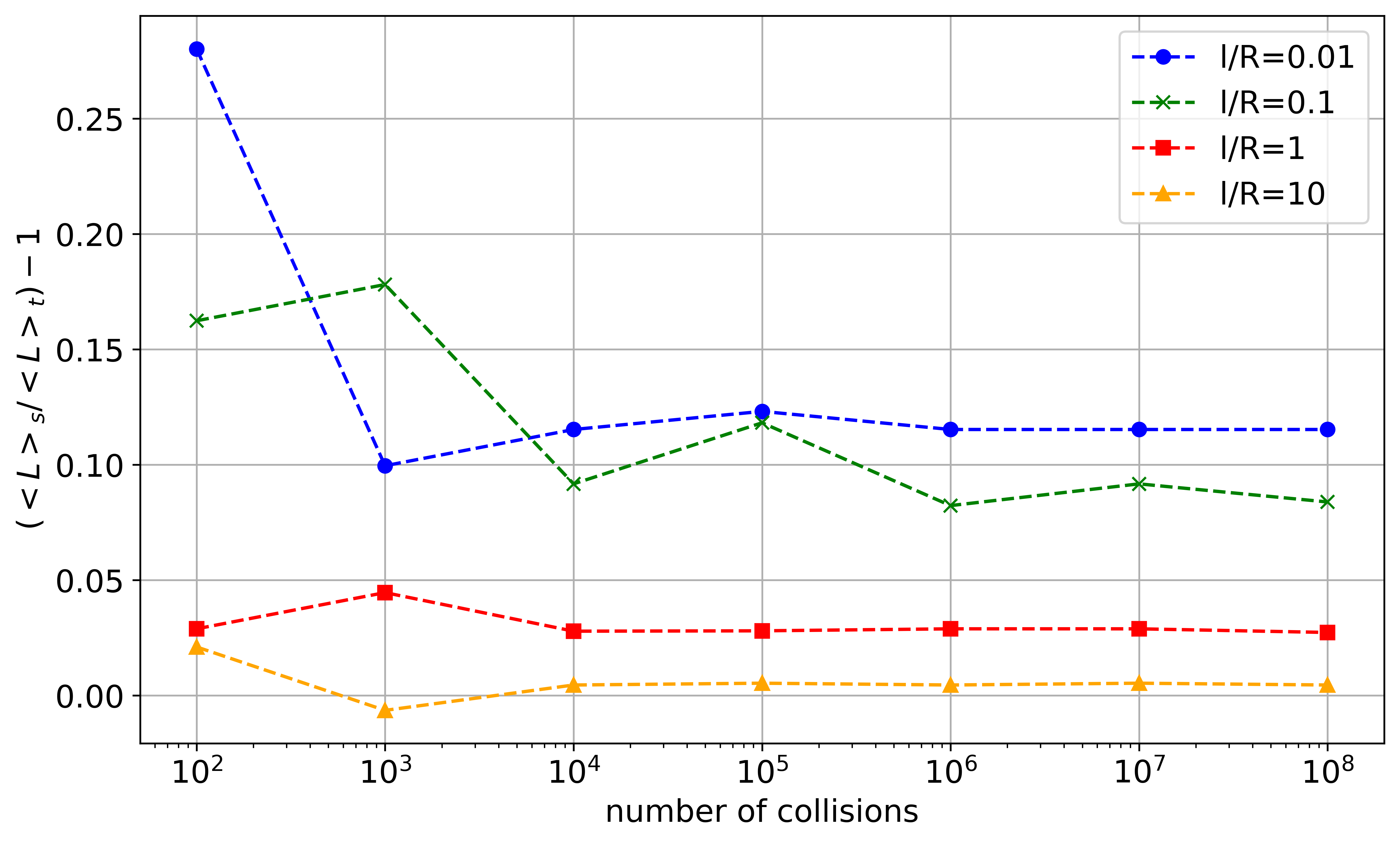}
        \caption{}
        \label{meanpathlengthabs}
    \end{subfigure}
    \caption{(a) Mean path length versus number of collisions with elastic boundaries. Each point is an average of 20 simulations. (b) Mean path length versus the number of collisions with random reflecting conditions at the boundary.}
\end{figure}

Let's now examine the scenario where the outgoing angle after a collision with the boundary is random. Initially, the absolute values of the mean path length under elastic boundary conditions, with an exponential step length distribution and uniformly random angle distribution, are presented in Fig. (\ref{meanpathlengthreflective}). These values approach the theoretical value of $\frac{\pi}{2}=1.57079...$ as the number of collisions approaches infinity. For contrast, the absolute values of the mean path length for the random reflecting angle condition are plotted in Fig. (\ref{meanpathlengthabs}). As can be seen, the invariant property no longer holds, as we anticipated in \cite{Artuso2024}. On the contrary, the mean path length does depend on $l/R$, i.e., the type of movement (diffusive or ballistic), as can be observed in Fig. (\ref{meanpathlengthabs}).

\begin{figure}[h!]
    \centering
    \includegraphics[width=\linewidth]{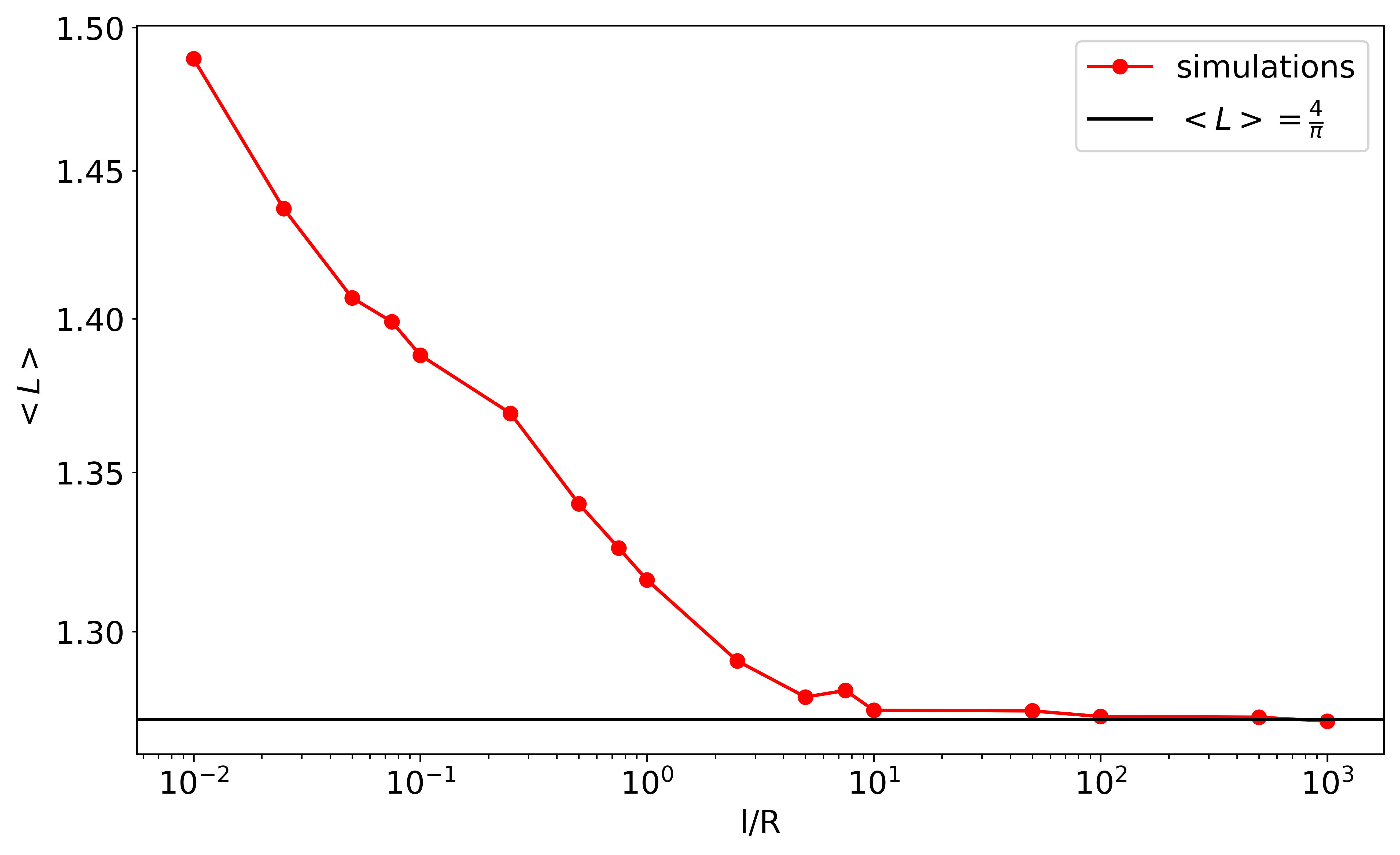}
    \caption{Mean path length as a function of $l/R$, $10^9$ collisions, random angle boundary conditions. The limit for $l/R\rightarrow\infty$ is the mean path length of a random billiard, $4/\pi=1.2732...$, represented by the black line.}
    \label{mplvslr}
\end{figure}

In Fig. (\ref{mplvslr}), we show this dependency in a $<L>$ versus $l/R$ plot for simulations with $10^9$ collisions. It can be seen that the mean path length tends to $4/\pi$, which is the mean path length of a random billiard as shown in \cite{Artuso2024}.
This behavior is consistent with the underlying physics of the system. In the limit where $l/R$ becomes very large—referred to as the ballistic regime—the particle's motion becomes dominated by long, straight trajectories between collisions with the boundary. In this regime, the particle travels in nearly straight lines across the domain, and its motion closely resembles that of a random billiard, where the particle's path is only altered by reflections at the boundary. For a random billiard, it has been shown that the mean path length, $<L>$, is exactly $4/\pi$. This convergence reflects the transition from diffusive behavior, characterized by shorter and more varied paths, to ballistic behavior, where the mean path length stabilizes at the expected value for a random billiard.

The MPLT holds true across various step length distributions—whether Gaussian, uniformly random, fixed, or exponentially distributed—as these do not affect the uniformity of particle incidence on the boundary, a critical assumption of the theorem. However, altering the angular distribution of the tumble events or the angle post-collision breaks this assumption. When the angular distribution deviates from being uniformly random, it introduces anisotropy or non-uniformity in the way particles interact with the boundary, violating the MPLT's conditions. For instance, fixed or Gaussian-distributed angles after tumble events create preferred directions of movement, leading to non-homogeneous wall collisions and thereby causing the theorem to fail. The MPLT does not hold under random boundary conditions, as opposed to elastic conditions, due to the way these conditions alter the uniformity and isotropy of particle interactions with the boundary of the domain. In contrast, under elastic boundary conditions, a particle's angle of reflection equals its angle of incidence, preserving the distribution of angles at which particles strike and leave the boundary. This uniformity ensures that the particle flux across the boundary maintains the assumptions required for the MPLT to apply.

\section{Distributions of path length between collisions}
\label{distributions:sec}

In the context of boundary interactions, homogeneous flux of particles at the boundary refers to a situation where particles strike the boundary uniformly, meaning that every point on the boundary has an equal likelihood of being hit by a particle. Isotropic flux means that particles approach the boundary from all directions with equal probability, ensuring that there is no preferred direction in the incidence of particles. After the evidence we showed in the last section, it is evident that under the condition of homogeneous and isotropic flux of particles at the boundary, the mean path length is determined by Cauchy's formula. However, this prompts the question: how does the distribution behave under different conditions? Let us delve into this inquiry.

\begin{figure}[h!]
    \centering
    \begin{subfigure}[b]{0.49\textwidth}
    	\centering
    	\includegraphics[width=\linewidth]{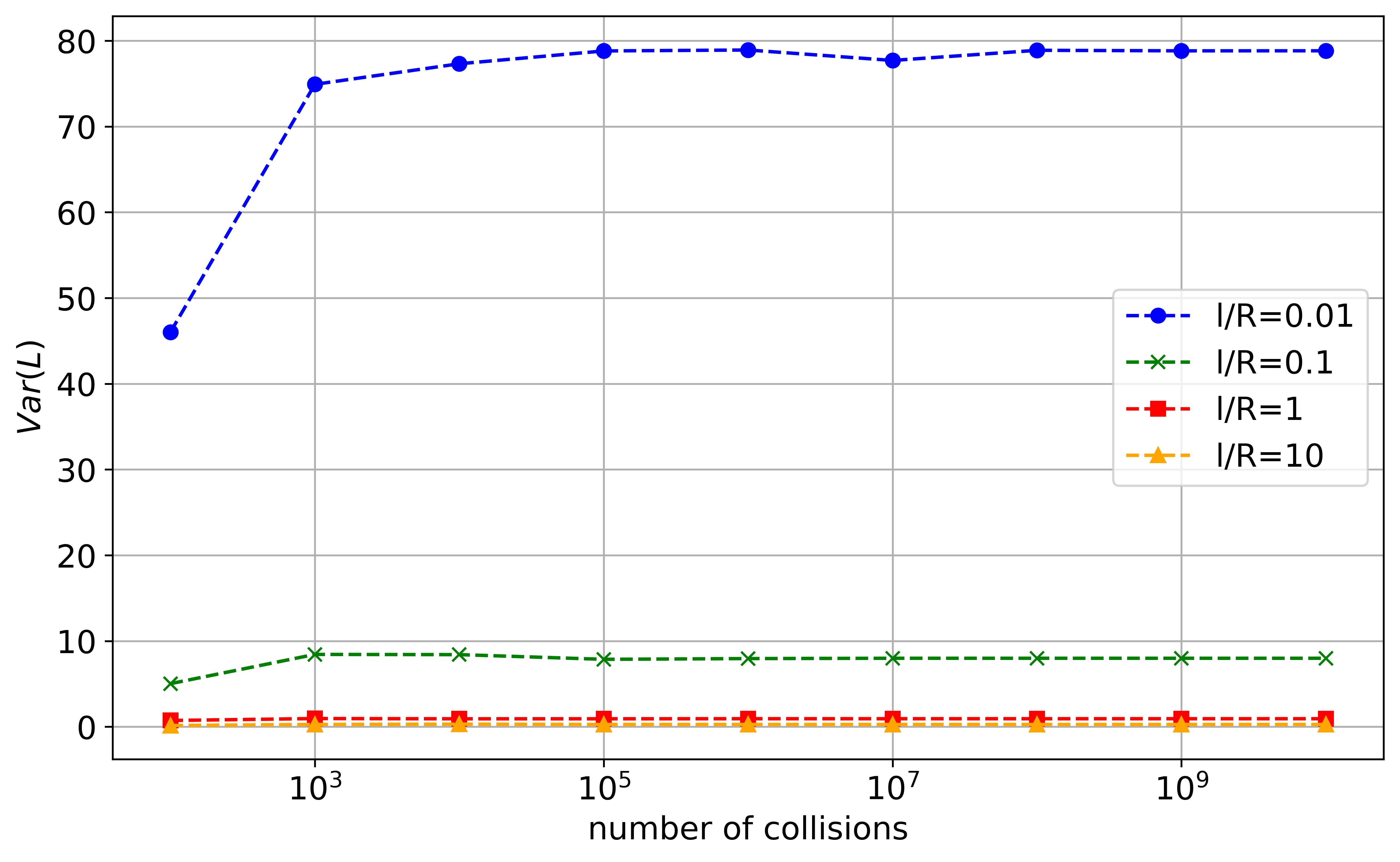}
	\caption{}
    \end{subfigure}
    \hfill
    \begin{subfigure}[b]{0.49\textwidth}
        \centering
        \includegraphics[width=\linewidth]{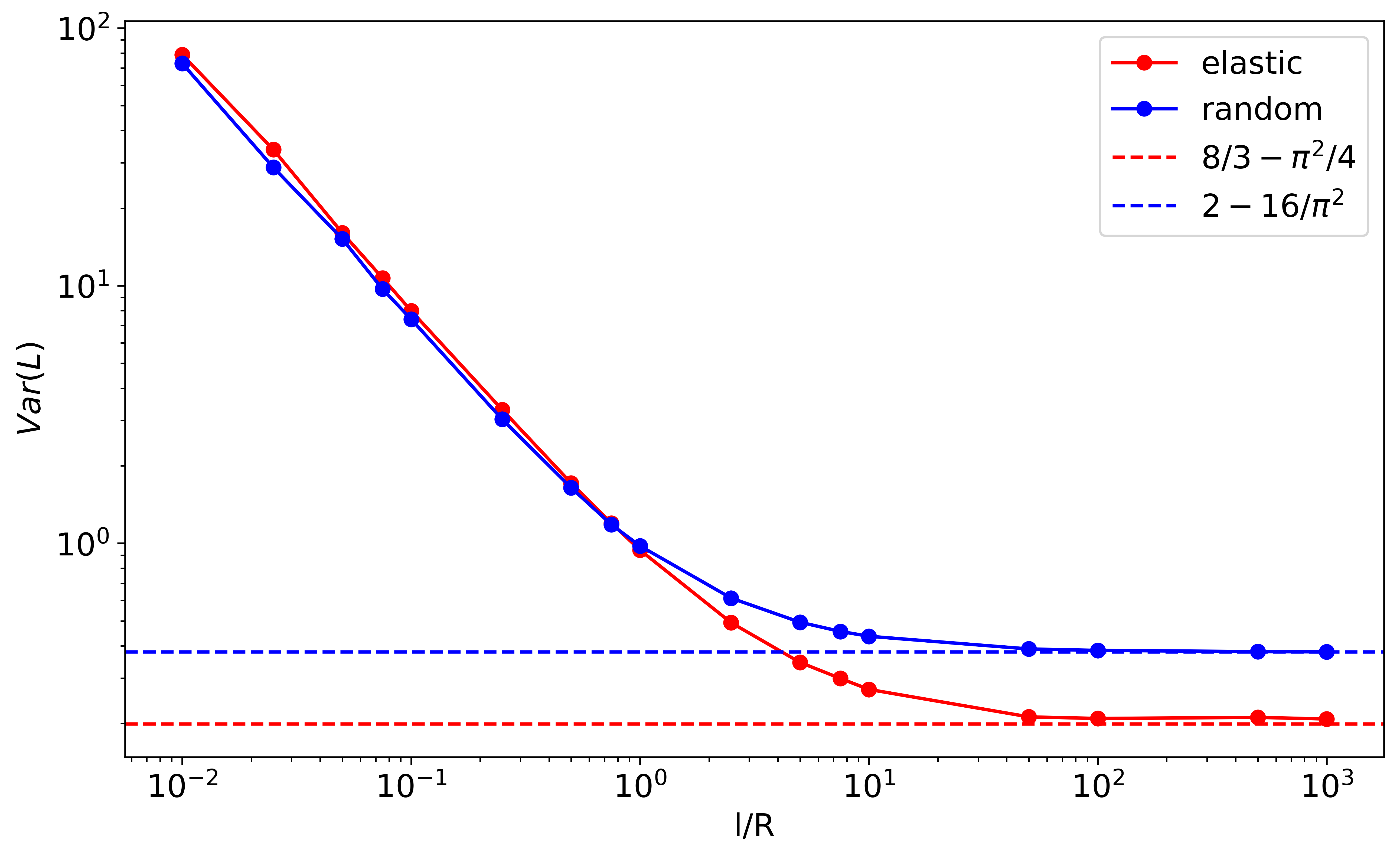}
        \caption{}
    \end{subfigure}
    \caption{(a) Variance of the path length versus the number of collisions with elastic reflecting conditions at the boundary. (b) Variance of the path length versus $l/R$, $10^9$ collisions. The dashed lines are the variances of elastic and random billiards.}
\label{variance}
\end{figure}

Hereinafter, we will focus on the particular case of exponential step length distribution and uniformly random angular distribution. In the subsequent analysis, we will further explore the dynamic properties of the path length distributions by examining the variance of these distributions. The variance provides crucial insight into the spread of the path length around its mean, which in this context is constant across all distributions at $<L>=\frac{\pi}{2}$ for the elastic boundary conditions, and variable as shown in Fig. (\ref{mplvslr}) for the random angle boundary condition. In Fig. (\ref{variance}), we plot the variance of the path length as a function of two parameters: the number of collisions for the elastic condition, and the parameter $l/R$ for both, elastic and random boundary conditions. This plot illustrates how the variance evolves as the motion transitions from diffusive to quasi-ballistic regimes across different scalings of $l/R$. The limit as $l/R\rightarrow\infty$ is $Var(L)_{elastic}=8/3-\pi^2/4\sim0.199$ for the elastic case. On the contrary, the random billiard has the limit $Var(L)_{random}=2-16/\pi^2\sim0.378$. These limits can be obtained by the following calculation.

Let first consider usual billiards. Usual billiards involve elastic (specular) reflections: the ingoing and outgoing angles coincide, or, equivalently

\begin{equation}
\label{vK}
v_1=v_0-2 (n(\tilde{q})\cdot v_0)n(\tilde{q}),
\end{equation}

\nd where $v_0$ and $v_1$ are the incoming and outgoing velocities, respectively, $\tilde{q}$ is the point of impact along the boundary, and $n(\tilde{q})$ is the (inward) normal to the boundary at the point $\tilde{q}$. Notice that elastic reflection, besides being dictated by elastic collision law with an infinite massive boundary segment, makes the billiard flow a Hamiltonian system \cite{ChLe,BCL}: in particular if $\phi$ denotes the ingoing angle, and $\psi=f(\phi)$ the outgoing angle (for a deterministic collision rule) there is a phase space contraction $\alpha$ at each collision event, with

\begin{equation}
\label{phase-c}
\alpha=-\log \left(\frac{\sin \psi}{\sin \phi} f'(\phi)\right).
\end{equation}

So, specular reflections make the billiard flow a Hamiltonian system, for which volume in the phase space is conserved. Even if the following considerations are valid in any dimension, we will consider billiard tables in a bounded region $\Omega$ of $\mathbb{R}^2$: due to conservation of kinetic energy the phase space is thus $\Gamma=S^1\times \Omega$, and the (invariant) Lebesgue measure is

\begin{equation}
\label{inv-f}
d\mu=C_{\mu}\cdot d\theta\,dx\,dy \qquad C_\mu =(2\pi V_{\Omega})^{-1},
\end{equation}

\nd where $(x,y)$ denote spatial coordinates in $\Omega$, and $\theta$ is the angle denoting the direction. This invariant measure defines, for any flow observable, a phase average, which coincides a.e. with time averages provided the flow is ergodic (incidentally this is far from being trivial for large classes of {\em convex} billiards). The connection to billiard maps (discrete dynamical system with phase space coordinates $\tilde{q},\phi$), is made clear by a change of coordinates in the space phase volume: if we fix a point $x,y,\theta$, then the backward flow uniquely determines $\tilde{q},\phi$ (coordinates for the billiard map), together with $\tau_-(x,y,\theta)$ (time from the boundary to the point $x,y,\theta$): so if we change $dx,dy$ into the product of a line element along the flow ($dt$) and another one perpendicular to it ($\cos(\phi)\,d\tilde{q}$) \cite{SinaiRMS,CherFAA,CherJ}, we get

\begin{equation}
\label{CC}
d\mu=C_{\mu}\cdot d\phi\,\cos(\phi)\,d\tilde{q}\, dt,
\end{equation}

\nd {where $d\tilde{q}$ is the Lebesque measure along the boundary $\partial \Omega$. This is associated to an invariant measure for the billiard map:

\begin{equation}
\label{inv-m}
d\nu=C_{\nu}\cdot d\tilde{q}\,\cos(\phi) d\phi\qquad C_{\nu}=(2\cdot S_{\partial \Omega})^{-1}.
\end{equation}

Moreover we can easily get Cauchy formula for billiards:

\begin{equation}
\label{Cbil}
1=\int_{\Omega}\,\int_{S_1}\, d\mu=\frac{C_{\mu}}{C_{\nu}} \int_{\partial \Omega}\,\int_{-\pi/2}^{\pi/2}\, d\nu\, \tau(\tilde{q},\phi),
\end{equation}

\nd namely, the phase average of the collision to collision time (averaged with the invariant measure for the billiard map), obeys (in dimension $d=2$)

\begin{equation}
\label{pCauchy}
\langle \tau \rangle_{\nu}=\frac{C_{\nu}}{C_{\mu}}=\pi \frac{V_{\Omega}}{S_{\partial \Omega}}.
\end{equation}

This identity establishes Cauchy law (for the average of collision to collision times) for {\em ergodic} billiards (irrespective of geometric properties of the shape of the billiard table). While in former considerations the cosine law (\ref{inv-m}) arises as the invariant measure of a fully deterministic system, it was assumed as the distribution of reflected gas molecules irrespective of their incident angle by Knudsen \cite{KG}. Knudsen law has been discussed as arising from microstructure of the billiard boundary in \cite{FY}, while \cite{KCel} investigate how it statistically arises from microscopic molecular simulations.

After these considerations, replacing the value of $C_\mu$ for the particular case of a circle in the cosine law, we obtain that, for the elastic case, the measure is $\mu(d\theta)=\frac{1}{2}\cos\theta d\theta$, and, by the geometry of the problem, a chord between two points on the boundary has length $L=2R\cos\theta$, see Fig. (\ref{chord}), where $\theta$ is the outgoing angle from the first point to the second one, measured with respect to the normal. Therefore,

\begin{equation}
\begin{split}
	<L^2>_{elastic}&=\int_{-\pi/2}^{\pi/2}(2R\cos\theta)^2\times\frac{1}{2}\cos\theta\times d\theta\\
	&=\frac{8}{3}R^2,
\end{split}
\end{equation}

\nd and the variance for our case $R=1$ is $Var(L)_{elastic}=<L^2>-<L>^2=8/3-\pi^2/4$. On the other hand, in the case of random boundary conditions, the measure is $\mu(d\theta)=d\theta/\pi$, corresponding to a collision rule $P_R(\theta) = 1/\pi$. The first moment is $<L>_{random}=4/\pi$ \cite{Artuso2024}, and the second moment is

\begin{equation}
	<L^2>_{random}=\int_{-\pi/2}^{\pi/2}(2R\cos\theta)^2\times\frac{1}{\pi}\times d\theta=2R^2.
\end{equation}

Therefore the variance (with $R=1$) is $Var(L)_{random}=2-16/\pi^2$.

\begin{figure*}[!]
    \centering
    \begin{subfigure}[b]{0.33\textwidth}
        \centering
        \includegraphics[width=\linewidth]{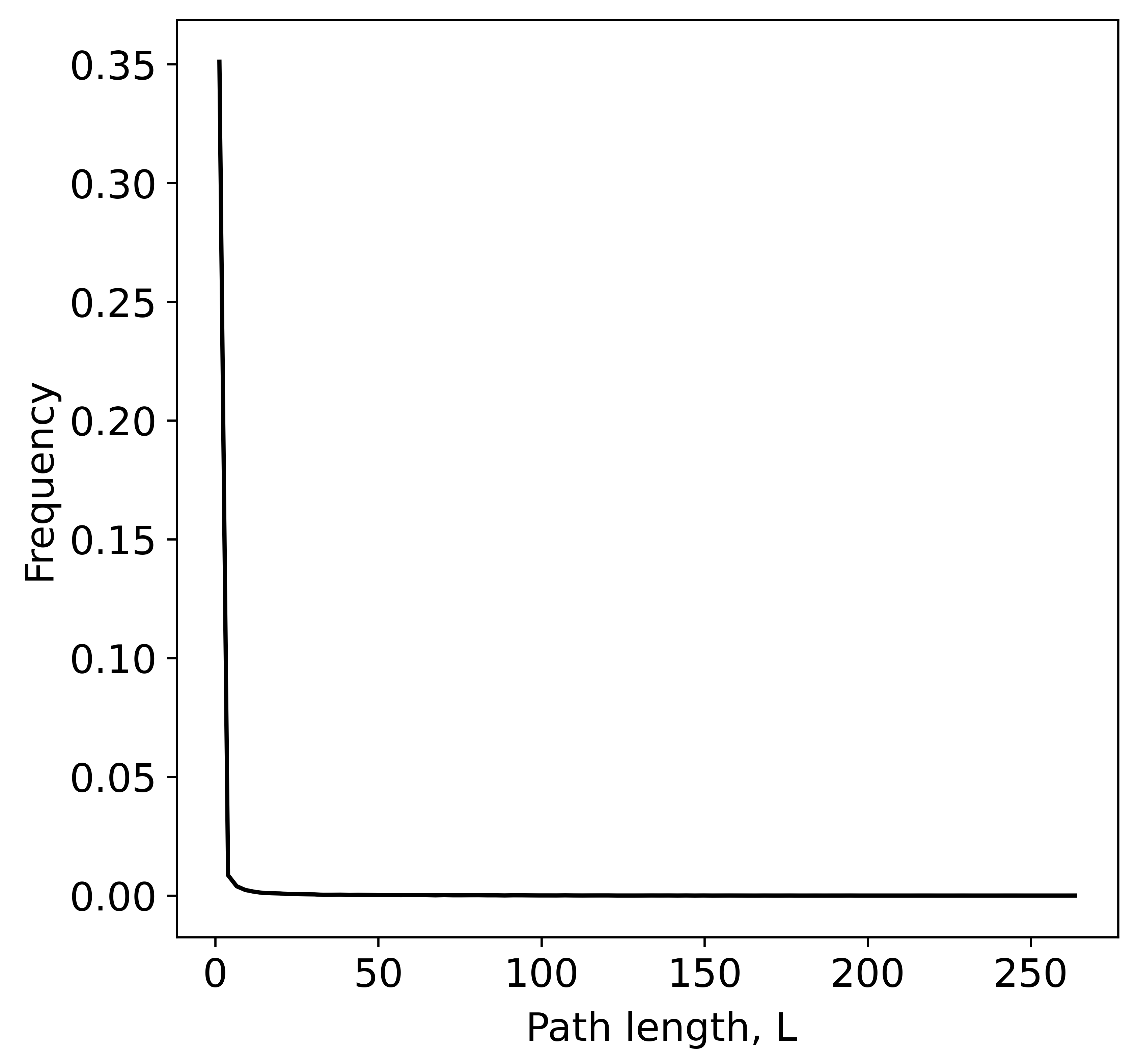}
        \caption{}
    \end{subfigure}
    \begin{subfigure}[b]{0.33\textwidth}
        \centering
        \includegraphics[width=\linewidth]{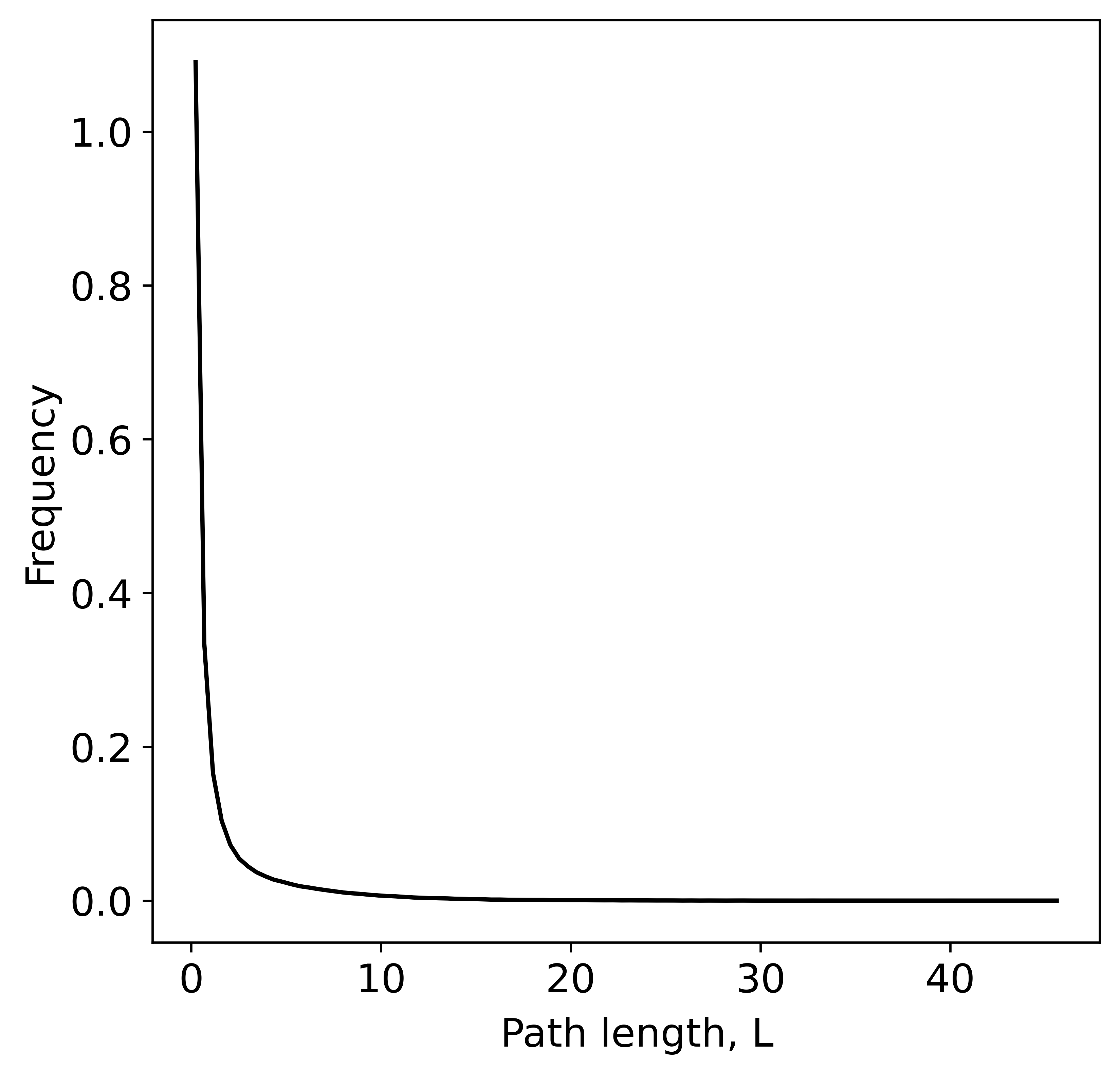}
        \caption{}
    \end{subfigure}
    \\
    \begin{subfigure}[b]{0.33\textwidth}
        \centering
        \includegraphics[width=\linewidth]{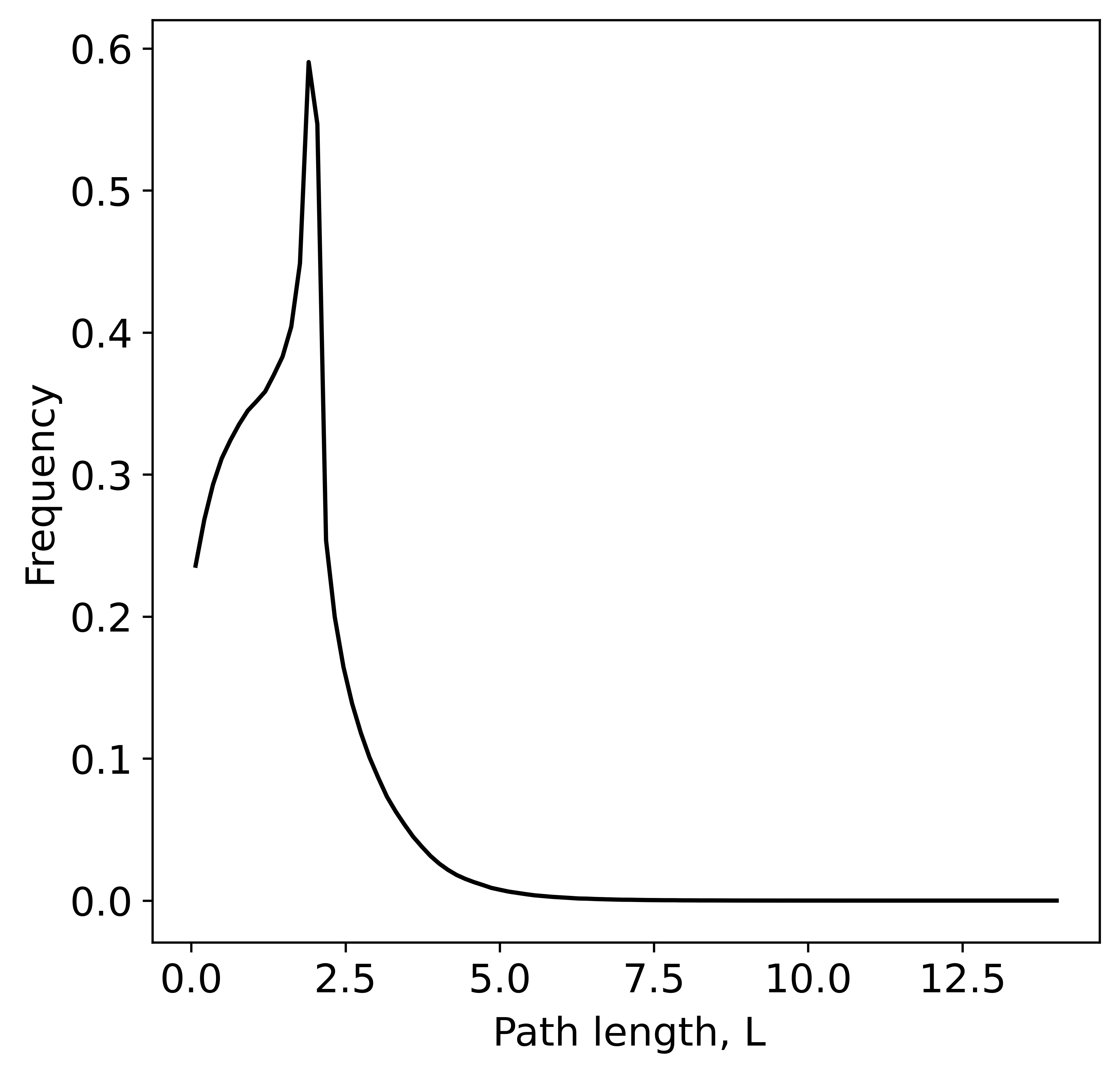}
        \caption{}
        \label{1MPL}
    \end{subfigure}
    \begin{subfigure}[b]{0.33\textwidth}
        \centering
        \includegraphics[width=\linewidth]{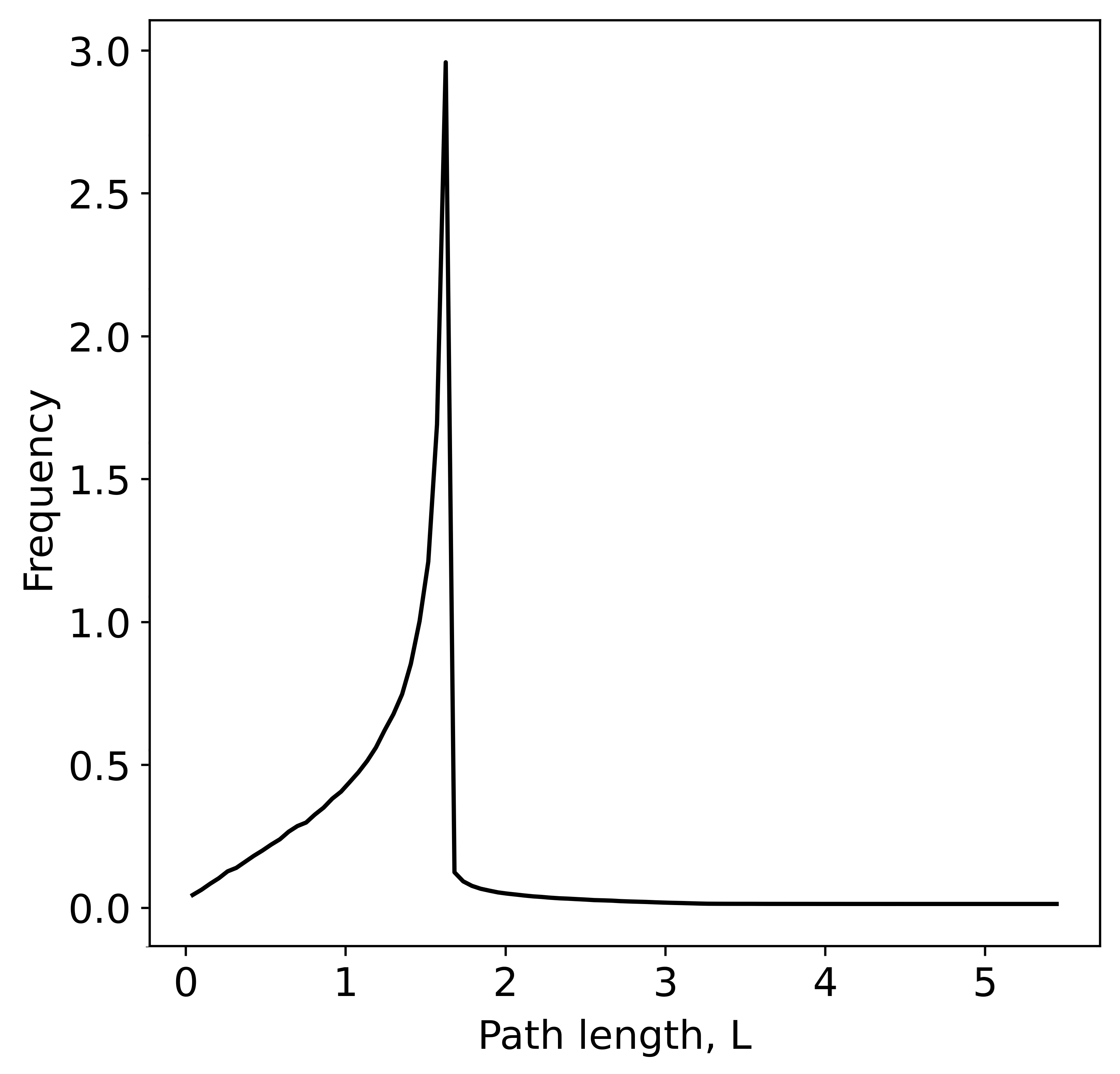}
        \caption{}
        \label{10MPL}
    \end{subfigure}
    \caption{Path length distribution with elastic boundaries, $10^7$ collisions, 100 bins. (a) $l/R=0.01$, (b) $l/R=0.1$, (c) $l/R=1$, (d) $l/R=10$.}
    \label{distributions}
\end{figure*}

A clearer picture can be obtained by examining the distributions. In Fig. (\ref{distributions}), the distribution of path lengths between collisions is shown, featuring elastic boundaries and the conditions previously mentioned, for various values of the parameter $l/R$: (a) $l/R=0.01$, (b) $l/R=0.1$, (c) $l/R=1$, (d) $l/R=10$. As illustrated in Fig. (\ref{meanpathlength}), while all distributions share the same mean, $<L>=\pi/2$, one can observe how the distribution transitions from diffusive (a) to quasi-ballistic (d). 

\begin{figure*}[!]
    \centering
    \begin{subfigure}[b]{0.32\textwidth}
        \centering
        \includegraphics[width=\linewidth]{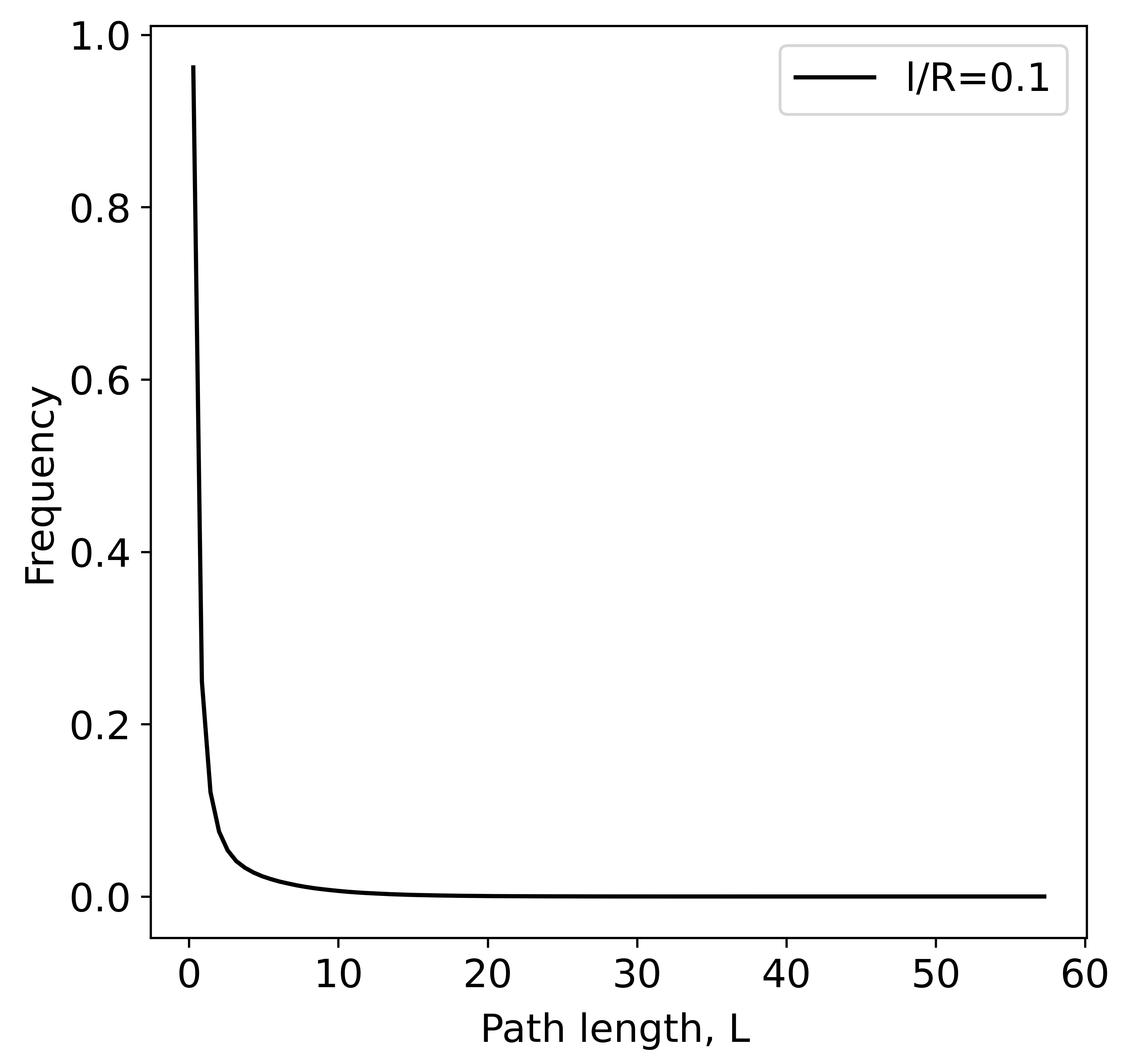}
        \caption{}
    \end{subfigure}
    \begin{subfigure}[b]{0.33\textwidth}
        \centering
        \includegraphics[width=\linewidth]{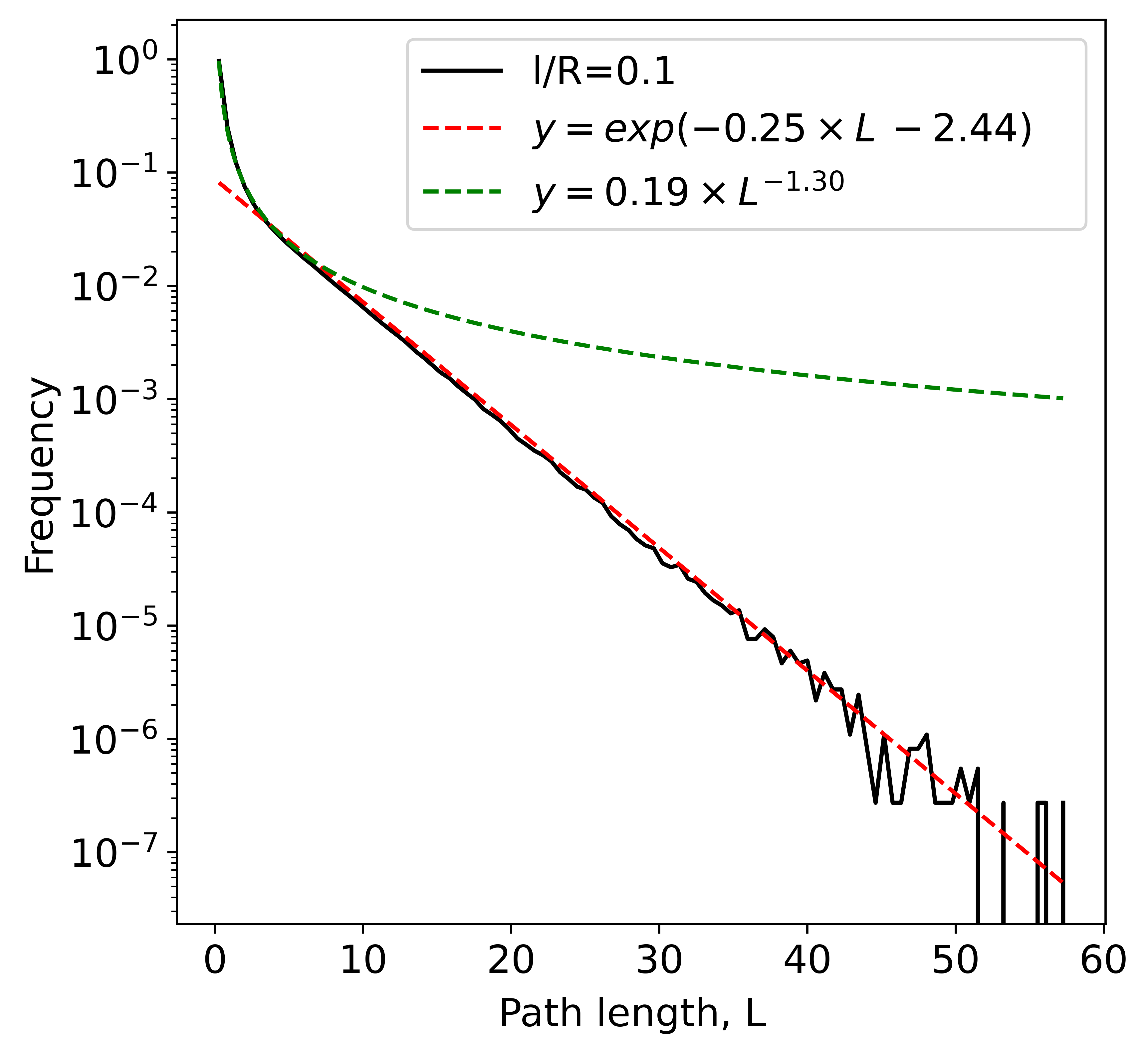}
        \caption{}
    \end{subfigure}
    \begin{subfigure}[b]{0.33\textwidth}
        \centering
        \includegraphics[width=\linewidth]{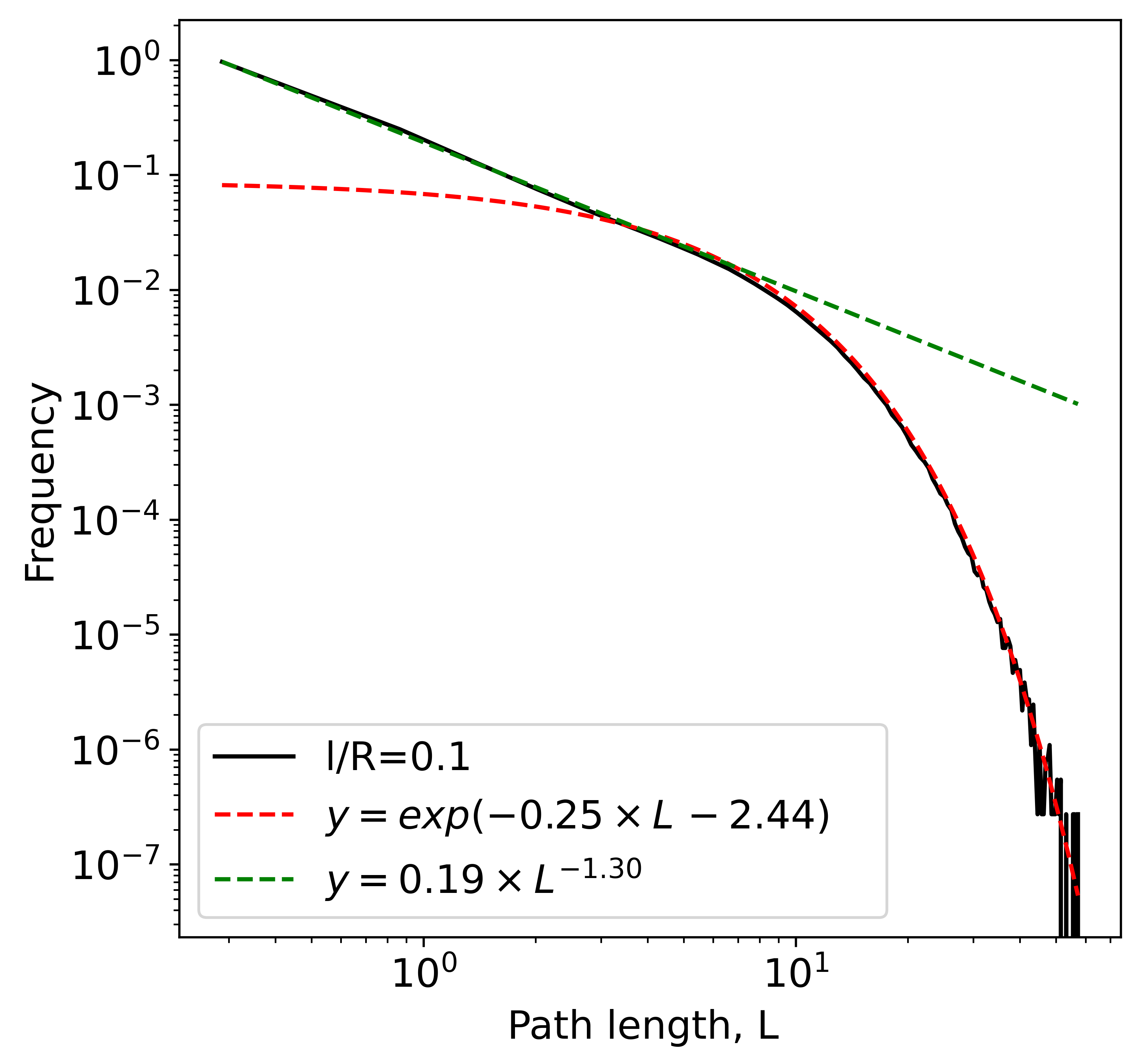}
        \caption{}
    \end{subfigure}
    \caption{Path length distribution with elastic boundaries, $10^7$ collisions, 100 bins. Ratio $l/R=0.1$. (a) linear, (b) semilog, and (c) log-log representation. Red lines are exponential fittings, green lines are power-law fittings.}
    \label{0.1}
\end{figure*}

\begin{figure*}[!]
    \centering
    \begin{subfigure}[b]{0.32\textwidth}
        \centering
        \includegraphics[width=\linewidth]{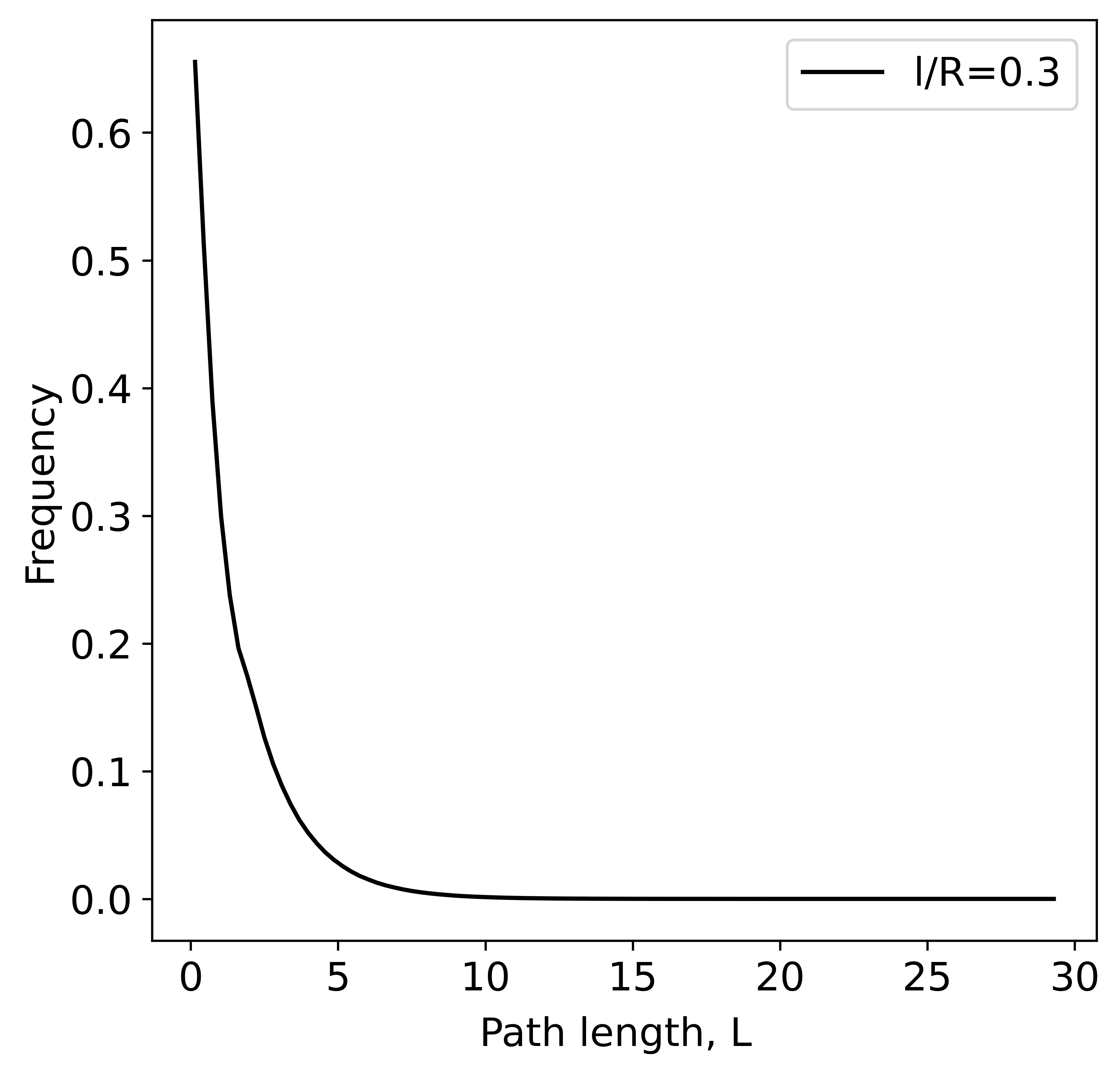}
        \caption{}
    \end{subfigure}
    \hfill
    \begin{subfigure}[b]{0.32\textwidth}
        \centering
        \includegraphics[width=\linewidth]{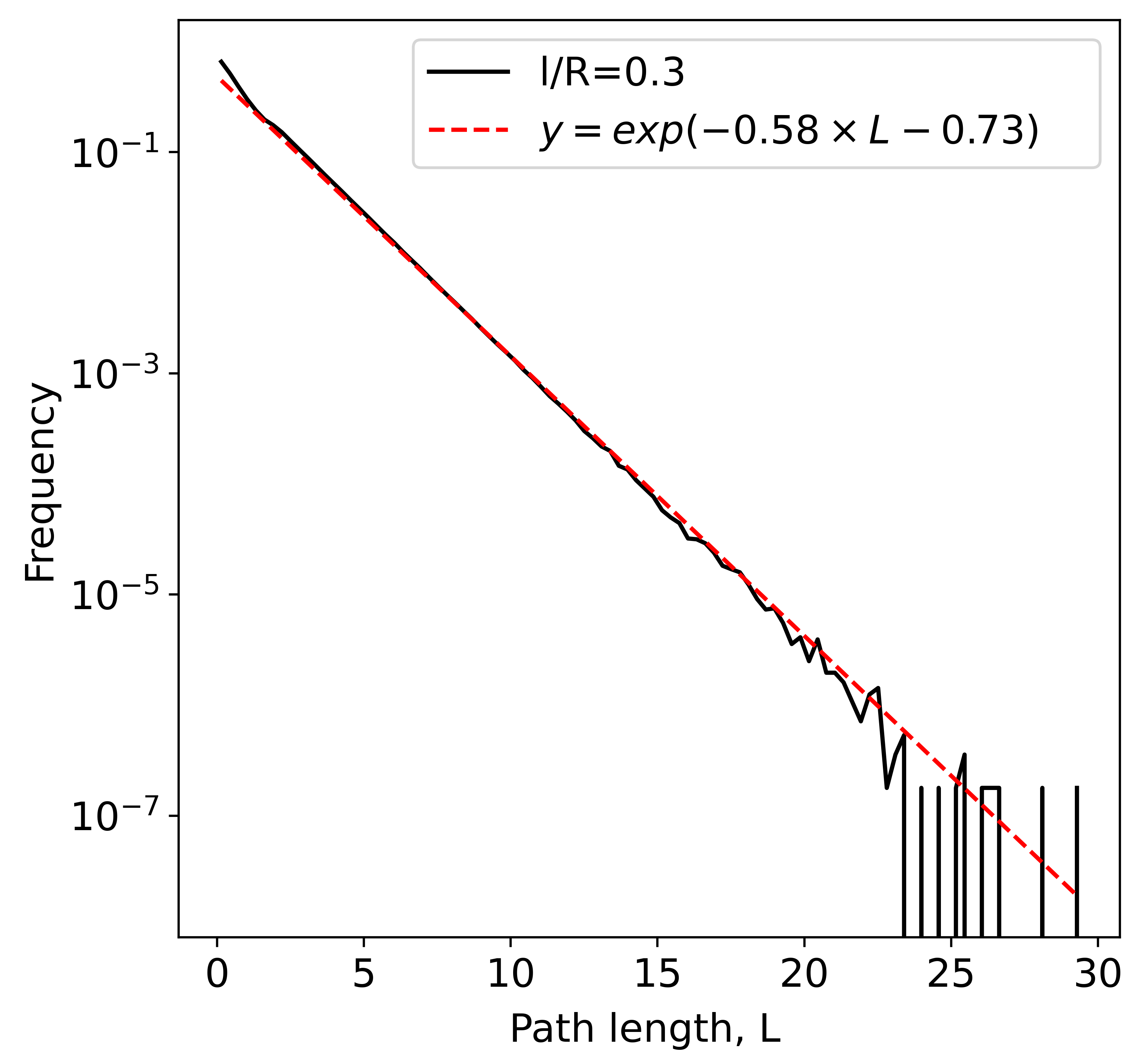}
        \caption{}
    \end{subfigure}
    \hfill
    \begin{subfigure}[b]{0.32\textwidth}
        \centering
        \includegraphics[width=\linewidth]{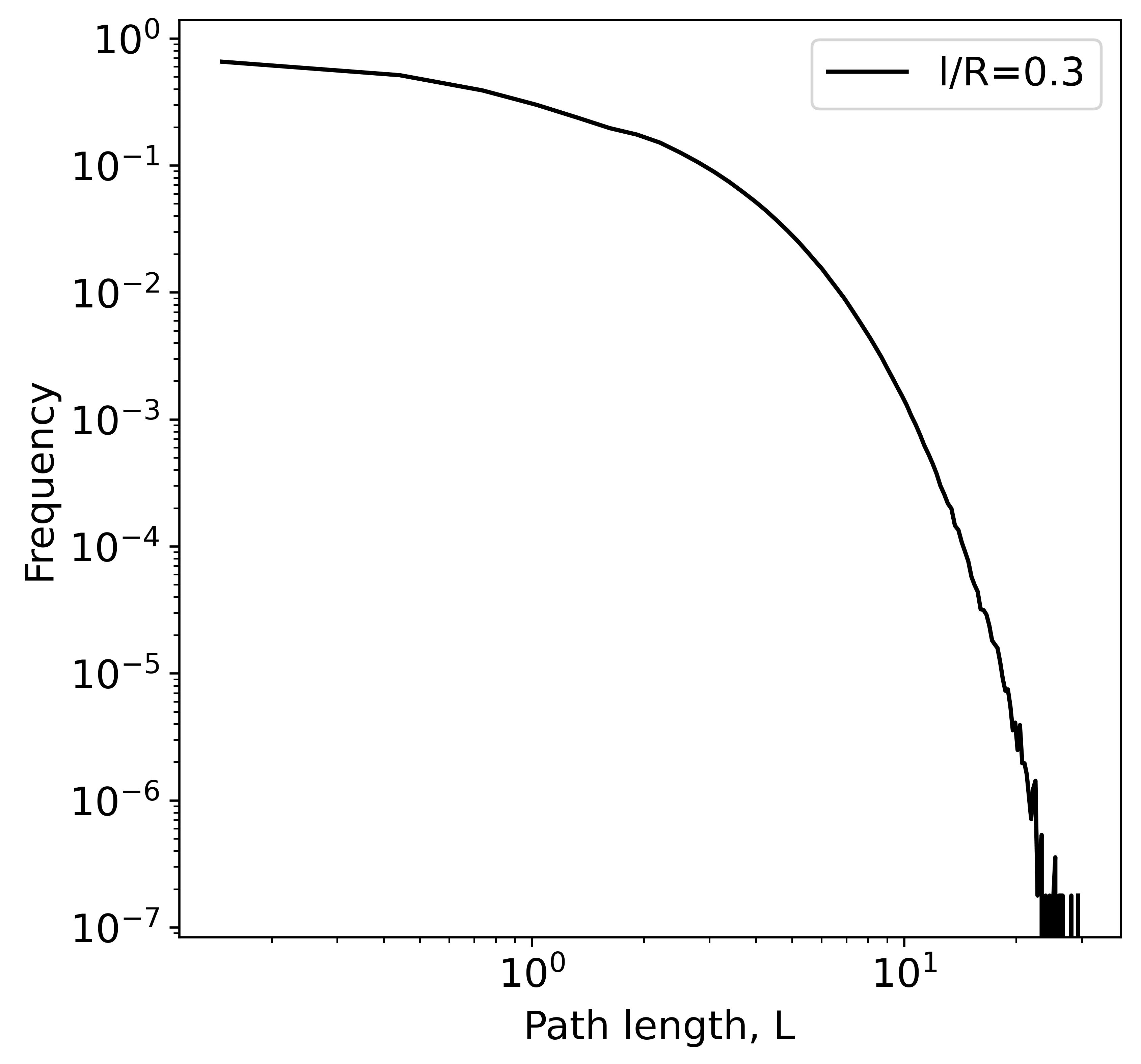}
        \caption{}
    \end{subfigure}
    \caption{Path length distribution with elastic boundaries, $10^7$ collisions, 100 bins. Ratio $l/R=0.3$. (a) linear, (b) semilog, and (c) log-log representation. The red line is an exponential fitting of the distribution.}
    \label{0.3}
\end{figure*}

Interestingly, when $l/R<1$, the distribution exhibits an exponential decay with a power-law deviation for small values of length, as illustrated in Figs. (\ref{0.1}) and (\ref{0.3}). The distributions are presented in (a) linear, (b) semilog, and (c) log-log formats. It is noteworthy that for $l/R=0.3$, the semilog representation displays a straight line, whereas for $l/R=0.05$, it is the log-log representation that appears straight. As we decrease the ratio $l/R$, the power-law deviation for small lengths becomes more pronounced. The decay parameter of the exponential fitting, $\gamma$, depends on $l/R$, as shown in Fig. (\ref{gamma}). A quadratic fitting seems to represent the numerical data quite well, but we have no theoretical argument to support this choice. The best values of the fit are presented in the label of Fig. (\ref{gamma}).

\begin{figure}[!]
    \centering
    \includegraphics[width=\linewidth]{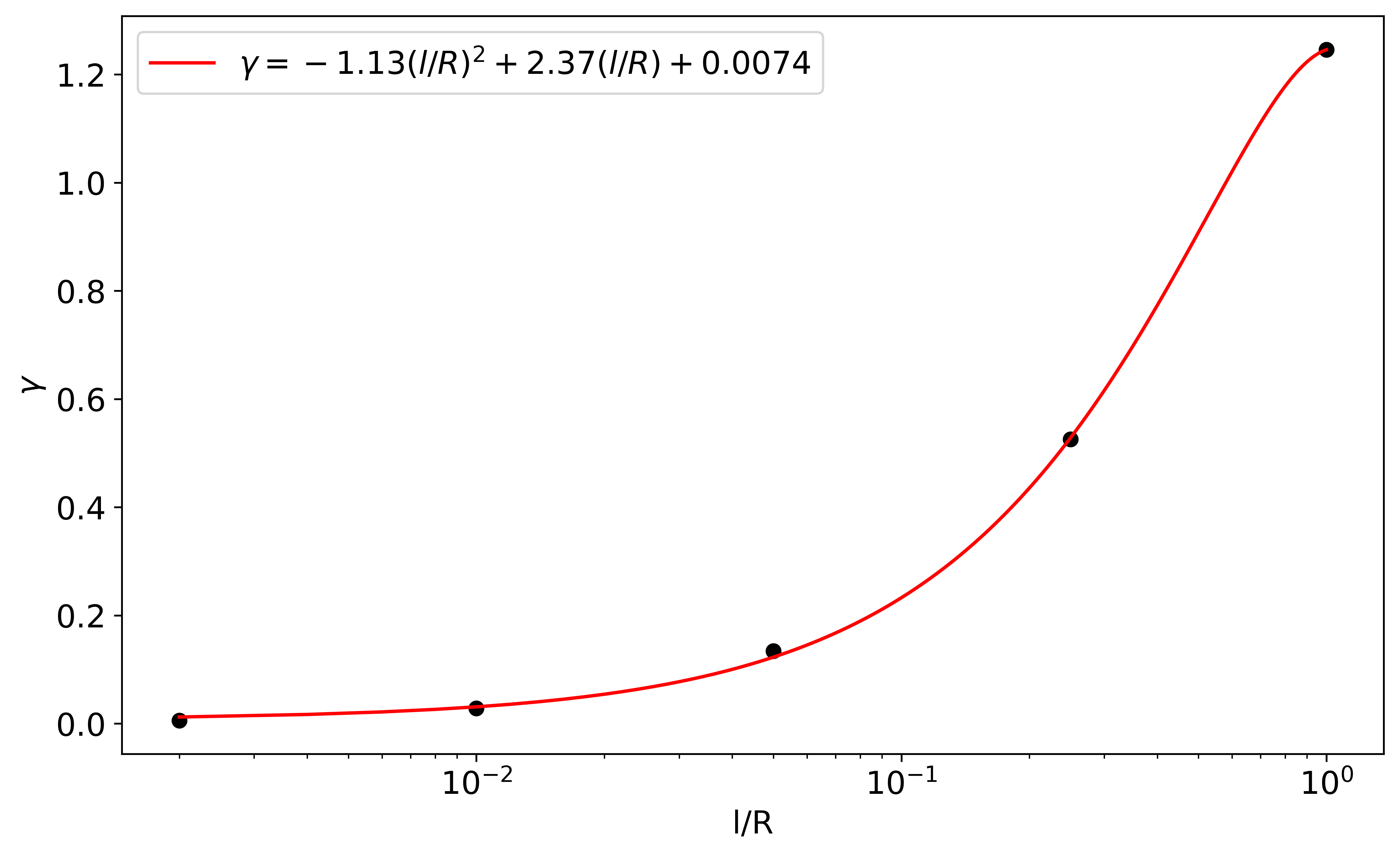}
    \caption{Decay parameter $\gamma$ of the exponential fitting as a function of $l/R$. The red line is a quadratic fit of the points.}
    \label{gamma}
\end{figure}

When $l/R>1$, the distributions, as shown in Fig. (\ref{10MPL}), for example, are a combination of diffusive and ballistic distributions. This occurs because sometimes the step length is equal to or greater than R, contributing to the ballistic mode, while at other times the step length is small and therefore contributes to the diffusive mode. Let us denote by $\tau$ the length of a chord between two points on the boundary. Recall that $\tau=2R\cos\theta$, see Fig. (\ref{chord}), where $\theta$ is the outgoing angle from the first point to the second, measured with respect to the normal. Then we have

\begin{equation}
	|\frac{d\tau}{d\theta}| = 2R\sin\theta = \sqrt{4R^2-\tau^2},\hspace{3mm} \theta\in[0,\pi/2].
\end{equation}

When the boundary conditions are elastic, the distribution verifies

\begin{equation}
	\Pi(\tau)d\tau = \cos\theta d\theta,
\end{equation}

\begin{equation}
	\Pi(\tau) = \frac{\cos\theta}{|d\tau/d\theta|} = \frac{\tau}{2R\sqrt{4R^2-\tau^2}}.
	\label{ballisticeq}
\end{equation}

As can be seen in Fig. (\ref{ballistic}), the ballistic part can be described with high accuracy by Eq. (\ref{ballisticeq}). In this figure, we show the ballistic part of the path length distribution for elastic boundary conditions, an exponential step length distribution, and a uniformly random angular distribution. The distribution is constructed with 100 bins and $10^7$ collisions, for the case $l/R=10$. 

\begin{figure}[!]
    \centering
    \begin{subfigure}[b]{0.33\textwidth}
    	\centering
  	\includegraphics[width=.75\linewidth]{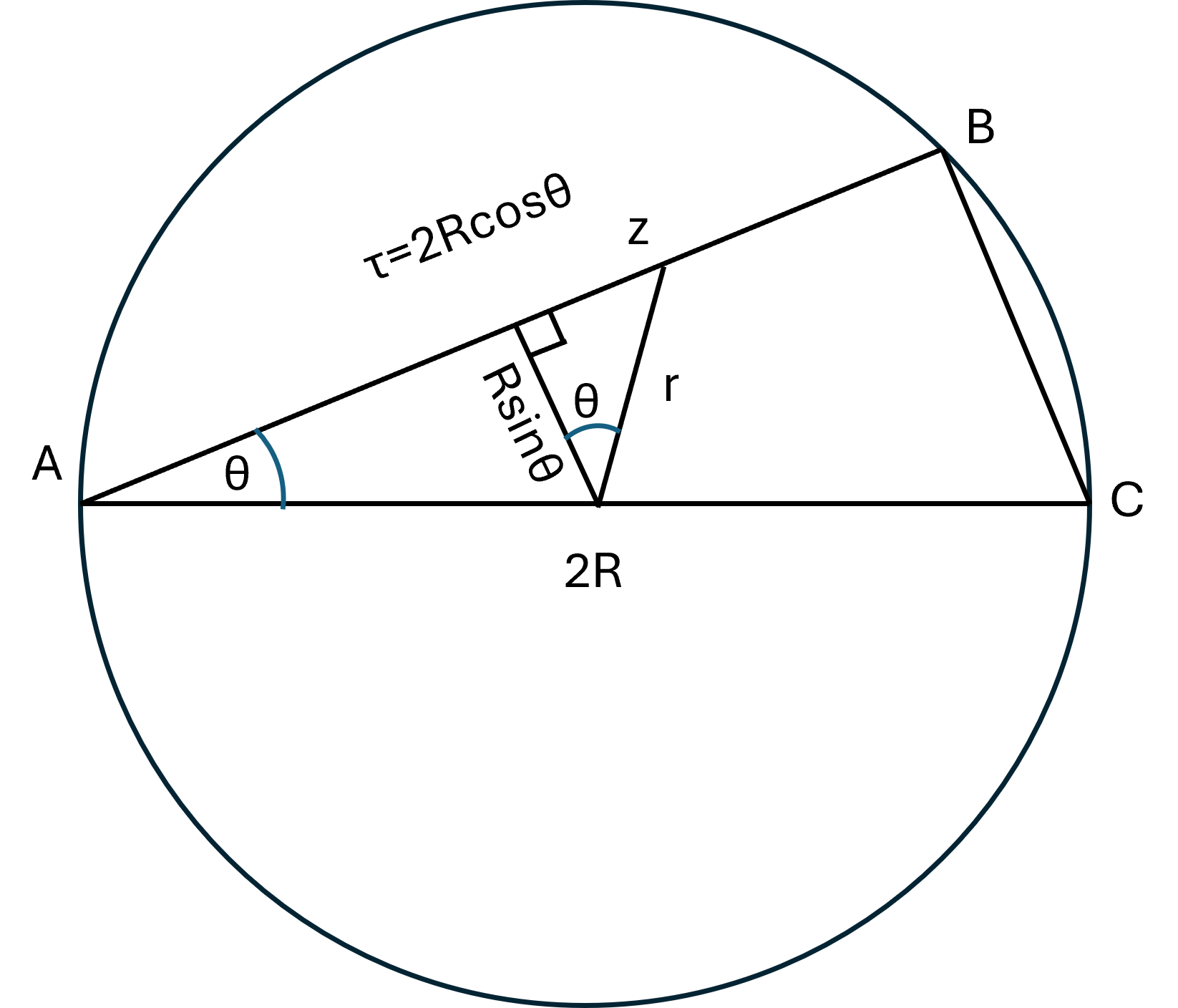}
	\caption{}
	\label{chord}
    \end{subfigure}
    \hfill
    \begin{subfigure}[b]{0.33\textwidth}
        \centering
        \includegraphics[width=\linewidth]{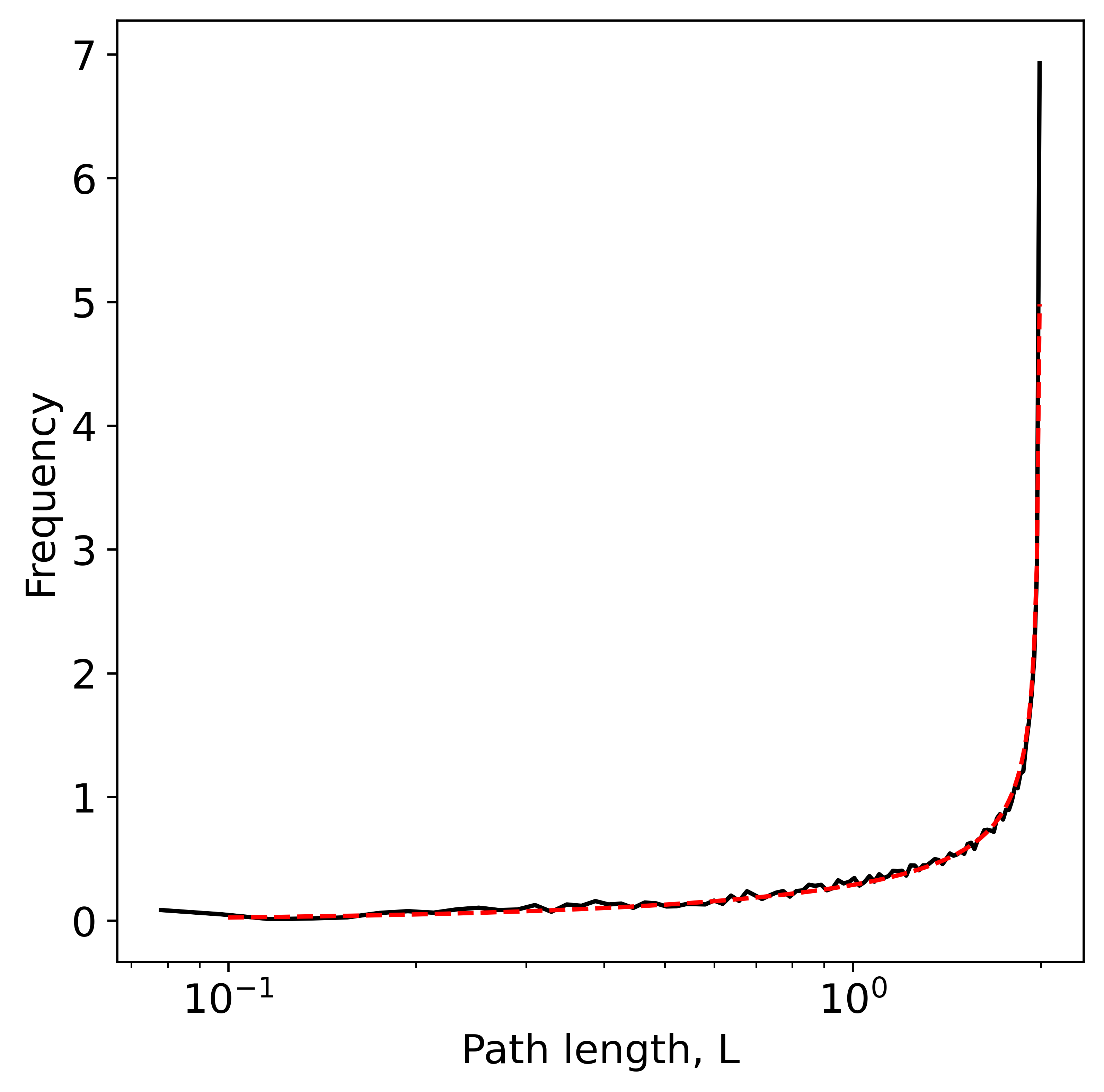}
        \caption{}
        \label{ballistic}
    \end{subfigure}
    \caption{(a) Geometry of a chord. (b) Ballistic part of the path length distribution with elastic boundaries. In black: simulation for $l/R = 10$, $10^7$ collisions, and 100 bins. In red: theoretical model, eq. (\ref{ballisticeq}).}
\end{figure}

\section{Radial distributions}
\label{radial:sec}

Let us now study the radial distributions, i.e.,  the probability of finding a particle at a given distance $r$ from the center of the confined domain, $p(r)$.This distribution helps in understanding how particles spread out within a bounded area, which is essential for predicting escape times, collision frequencies, and interaction rates with boundaries. The radial distribution of run-and-tumble particles in a confined two-dimensional domain is a critical aspect that provides insight into the spatial dynamics and density distribution of particles within the domain. This section explores the relevance of radial distribution in understanding run-and-tumble movement.

We performed the simulations with an exponential step length distribution and a uniformly random angular distribution. The radial distribution was found to be linear under elastic boundary conditions for all values of the ratio $l/R$. For instance, in Fig. (\ref{radialprob}), we show the distribution for the case $l/R=0.3$. This can be analytically calculated as follows. Imagine a chord between points at the boundary representing part of the trajectory of the particle. This is represented by the segment AB in Fig. (\ref{chord}). The probability, $W$, of being between a given point on the chord, $z$, and $z+dz$, is equal to the ratio between $dz$ and the length of the whole chord, $\tau$.

\begin{equation}
	W_{\theta}(r) dr= \frac{dz(r)}{\tau},
\end{equation}

\nd where $\theta$ is the angle measured with respect to the normal at the point A. Looking at the triangle ABC and taking into account that the segment AC is the diameter of the circle, we get that the length of the chord AB is $\tau=2R\cos\theta$. Therefore,

\begin{equation}
	W_{\theta}(r) dr= \frac{dz(r)}{2R\cos\theta},
\end{equation}

\nd where $r$ is the distance from the center of the circle to $z$, see Fig. (\ref{chord}). By geometry of the interior triangle we get

\begin{equation}
	r^2=z^2+R^2\sin^2\theta,
\end{equation}

\nd and therefore,

\begin{equation}
	\frac{dz}{dr}=\frac{2r}{\sqrt{r^2-R^2\sin^2\theta}},
\end{equation}

\begin{equation}
W_{\theta}(r) = \frac{1}{R \cos\theta}\frac{r}{\sqrt{r^2-R^2 \sin^2\theta}}.
\end{equation}

Taking into account that the measure is $1/2.\cos\theta.d\theta$ and the chord length is $\tau=2R\cos\theta$, we get that

\begin{equation}
<\tau> = R \int_{-\pi/2}^{\pi/2} cos^2 \theta d\theta = \frac{\pi R}{2}.
\end{equation}

Suppose you reconstruct the radial density by using the dynamics with discretized steps: then short segments
contribute less points than long segments: this can be taken into account exactly by a $\frac{\tau}{<\tau>}$ factor, and therefore, the radial probability can be calculated as

\begin{equation}
\begin{split}
	p_{elastic}(r) &= \int \frac{\tau}{<\tau>}W_\theta(r) \cos\theta d\theta\\
	&= \frac{4r}{\pi R} \int_0^{\arcsin(r/R)} \frac{\cos\theta d\theta}{\sqrt{r^2-R^2 \sin^2\theta}}.
\end{split}
\end{equation}

Making a change of variables,

\begin{equation}
\sin \phi = \frac{R}{r} \sin \theta
\end{equation}

\begin{equation}
\cos\phi d\phi = \frac{R}{r} \cos\theta d\theta
\end{equation}

We obtain,

\begin{equation}
p_{elastic}(r) = \frac{4r}{\pi R} \int_0^{\pi/2} \frac{\cos \phi d\phi}{r \cos\phi} = \frac{2r}{R^2}.
\label{radialeq}
\end{equation}

The last equation implies a uniform invariant density, $\rho$, since

\begin{equation}
	\rho_{elastic}= \frac{p_{elastic}(r)}{2\pi r} = \frac{1}{\pi R^2}.
\end{equation}

This theoretical model matches well with the simulation results, confirming the uniform distribution of particles when the boundary conditions are elastic.

\begin{figure}[!]
    \centering
    \begin{subfigure}[b]{0.50\textwidth}
    	\centering
  	\includegraphics[width=0.7\linewidth]{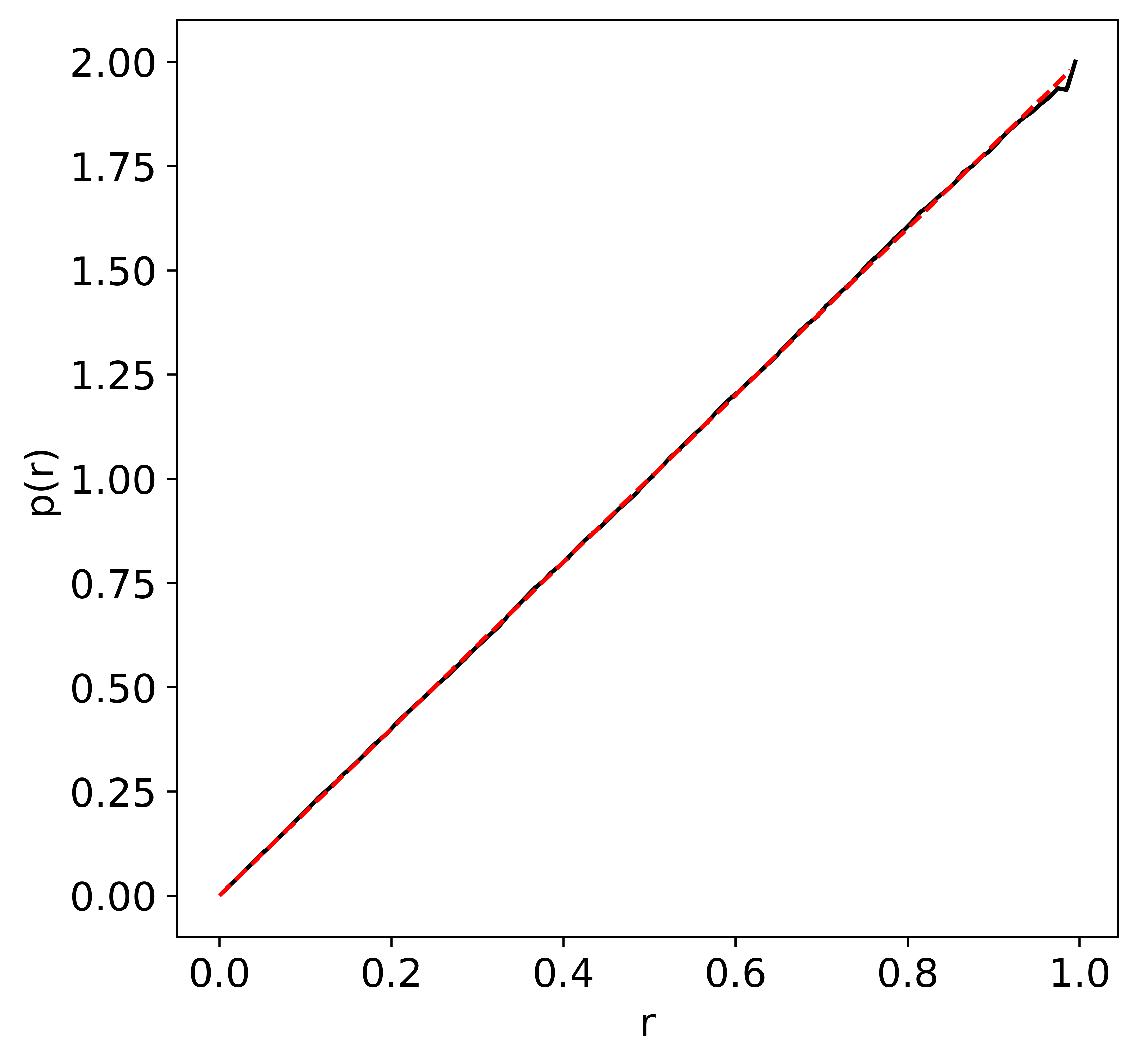}
	\caption{}
	\label{radialprob}
    \end{subfigure}
    \hfill
    \begin{subfigure}[b]{0.48\textwidth}
        \centering
        \includegraphics[width=0.7\linewidth]{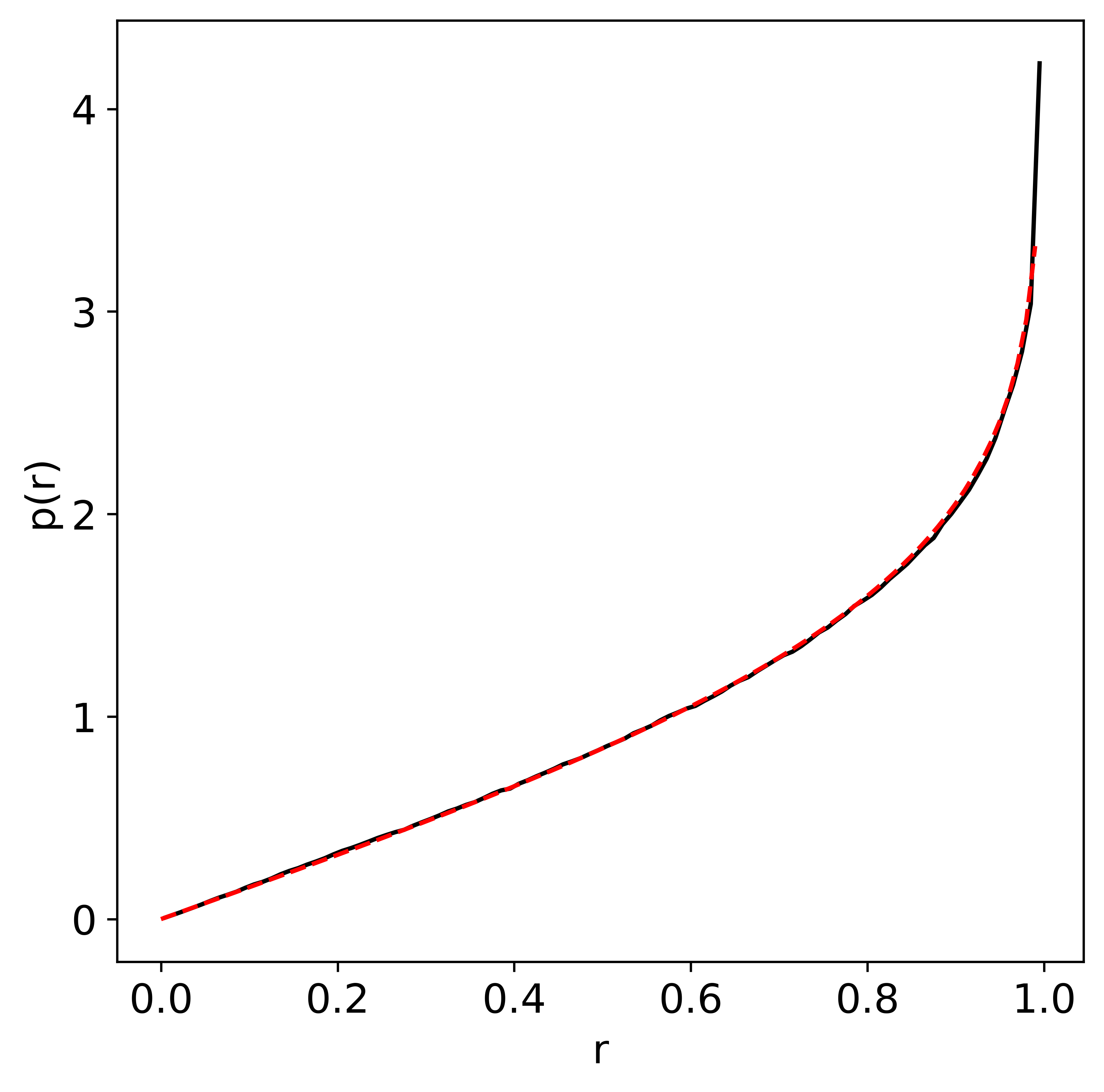}
        \caption{}
        \label{radialprob-random}
    \end{subfigure}
    \caption{(a) Radial distribution elastic boundary conditions, exponential step length and uniformly random angular distribution. In black: simulation for $l/R = 0.3$, $10^7$ collisions, and 100 bins. In red: theoretical model, eq. (\ref{radialeq}). The linear dependence indicates that there is a uniform invariante density. (b) Radial probability distribution for random reflecting angle at the boundary. In black: simulation for $l/R = 10$, $10^7$ collisions, 100 bins. In red: theoretical model, eq. (\ref{randomeq}).}
\end{figure}

In contrast, for random angle boundary conditions, the radial distribution deviates from linearity and is described by the complete elliptic integral of the first kind, which can be calculated as follows. In the case of random angle boundary conditions, the measure is $d\theta/\pi$, and therefore, 

\begin{equation}
	<\tau> = \frac{2R}{\pi} \int_{-\pi/2}^{\pi/2} \cos\theta d\theta = \frac{4R}{\pi}.
\end{equation}

The radial distribution is

\begin{equation}
\begin{split}
	p_{random}(r) &= \int \frac{\tau}{<\tau>}W_\theta(r) \frac{2}{\pi} d\theta\\
	&= \frac{r}{R}\int_0 ^{arcsin (r/R)} \frac{d\theta}{\sqrt{r^2-R^2\sin^2\theta}}.
\end{split}
\end{equation}

Using the same change of variable as before we get,

\begin{equation}
p_{random}(r) = \frac{r}{R^2}\int_0^{\pi/2} \frac{d\theta}{ \sqrt{1-\frac{r^2}{R^2}\sin^2\theta}}=\frac{r}{R^2} K\left(\frac{r}{R}\right),
\label{randomeq}
\end{equation}

\nd where $K(x)$ is the complete elliptic integral of the first kind, defined as \cite{Gradshteyn2007}

\begin{equation}
K(x) = \int_0^{\pi/2}\frac{ds}{\sqrt{1-x^2\sin^2s}}.
\end{equation}

This expression is compared with the simulation in Fig. (\ref{radialprob-random}) for the case $l/R=10$, and as can be seen, the agreement is satisfactory. An alternative procedure for calculating these distributions is presented in the appendix. This distribution shows that particles are more likely to be found near the walls of the domain under random angle reflections. These results seem to be in agreement with experiments with microbes, in which the microbes spend the majority of their time close to the walls of the domain \cite{Souzy2022, Ostapenko2018}. However, the experimental observation might have other physical explanations, such as hydrodynamical interactions, or other kind of microbe-wall interactions. The assumption that the interaction of a swimming microbe with a wall may be described as a collision could be subject to criticism and more realistic modelization of the interactions could be introduced. In fact, the study of the radial distribution can provide information about the mobility mechanism in experimental settings, since different distributions are generated depending on how the active agent interacts with the walls and, as we will show in the next section, the reorientation mechanism. It would be interesting, therefore, to see more experimental outcomes on this matter. A related concept, the boundary local time—i.e., how long a diffusing particle spends in the close vicinity of a confining boundary—thus plays a crucial role in the description of various surface-mediated phenomena, such as heterogeneous catalysis, permeation through semipermeable membranes, or surface relaxation in nuclear magnetic resonance \cite{Grebenkov2019}.

\section{Processes with memory}
\label{nemo:sec}

Let us now focus on a different, non-trivial case. Process memory refers to the phenomenon where the future state of a particle's motion is influenced by its past states, leading to non-Markovian dynamics. The concept of process memory plays a crucial role in understanding the reorientation mechanism of active agents in run-and-tumble movement. For example, some processes, like chemotaxis, introduce a kind of memory—namely, the ability of the particle to compare the concentration or gradient of a chemical across different steps of its movement.

\begin{figure}[!]
    \centering
    \begin{subfigure}[b]{0.49\textwidth}
    	\centering
    	\includegraphics[width=\linewidth]{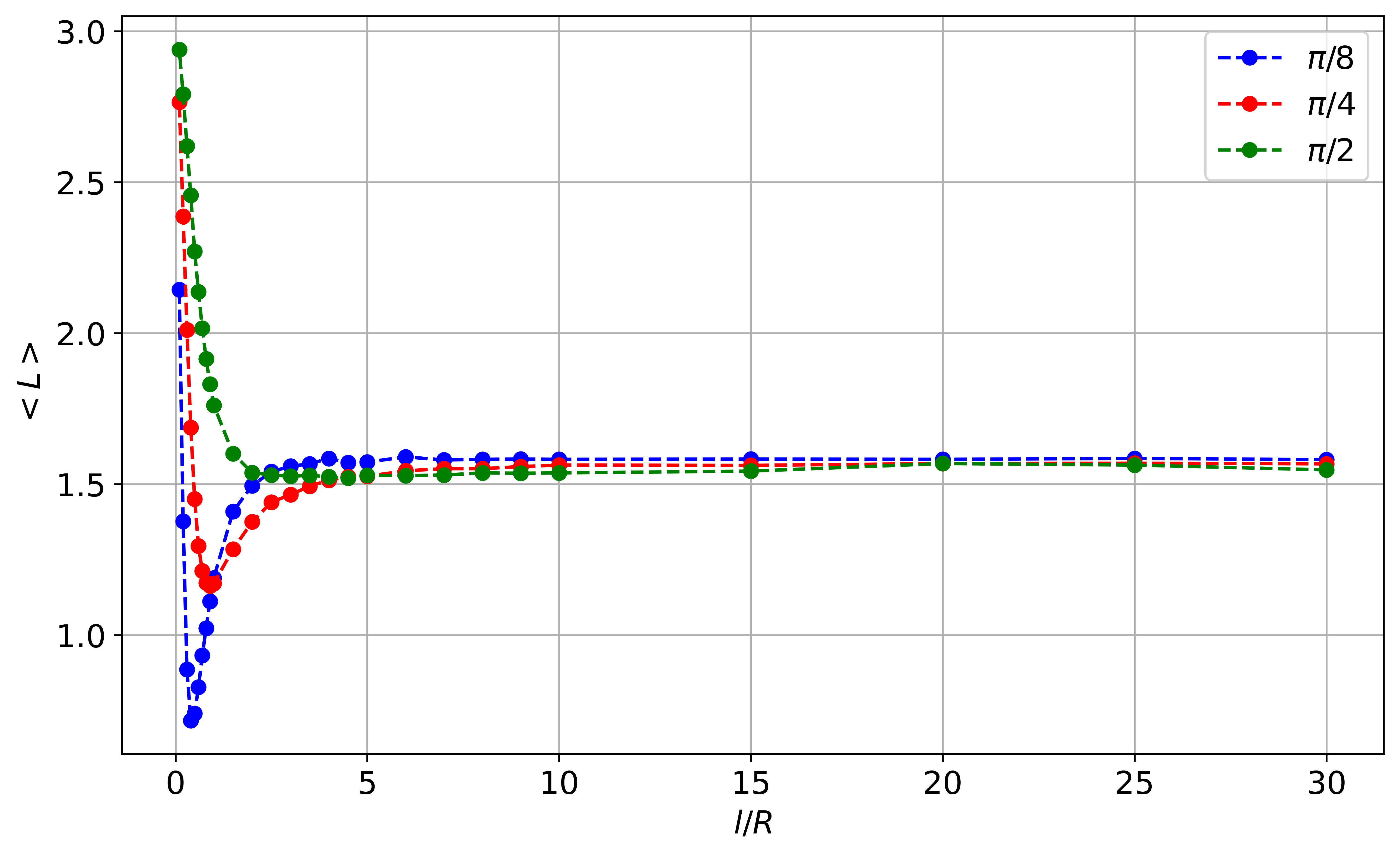}
	\caption{}
    \end{subfigure}
    \hfill
    \begin{subfigure}[b]{0.49\textwidth}
        \centering
        \includegraphics[width=\linewidth]{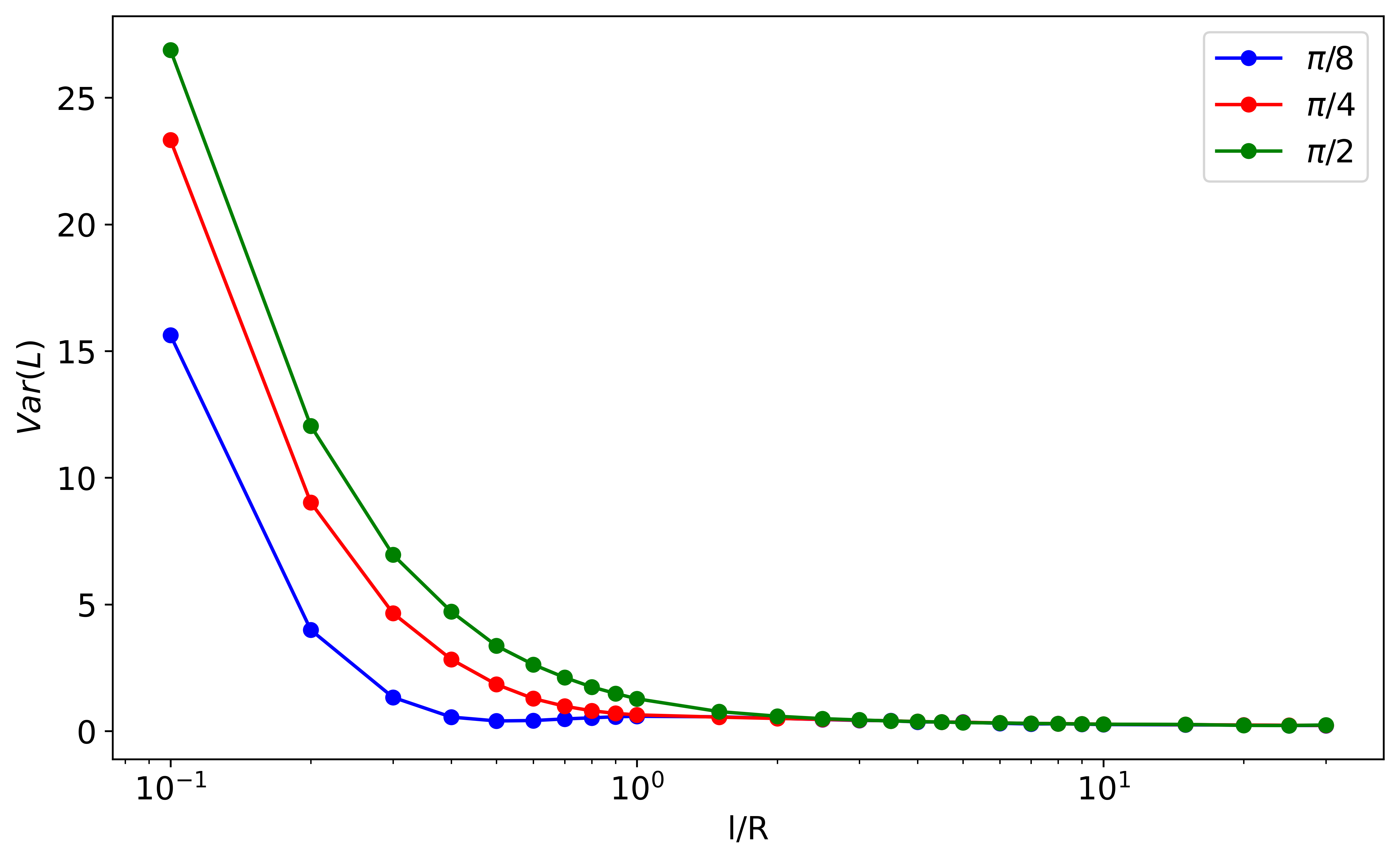}
        \caption{}
    \end{subfigure}
    \caption{Mean path length (a) and variance of path length (b) versus $l/R$ for three values of constant drift in the reorientation angle for simulations with $10^9$ collisions, elastic boundary, and exponential step length distribution.}
\label{NEMO}
\end{figure}

Process memory can significantly alter the motion patterns of particles, leading to deviations from theoretical predictions like the MPLT. We will introduce a simple memory process in the simulations, following this rule: the new reorientation angle of the tumble events is equal to the last one plus a constant, $\theta_n=\theta_{n-1}+C$. The step length distribution is exponential, and the boundary is elastic. We explore the mean and variance of the distributions for three different values of the constant $C$, as shown in Fig. (\ref{NEMO}).

\begin{figure*}[!]
    \centering
    \begin{subfigure}[b]{0.33\textwidth}
        \centering
        \includegraphics[width=\linewidth]{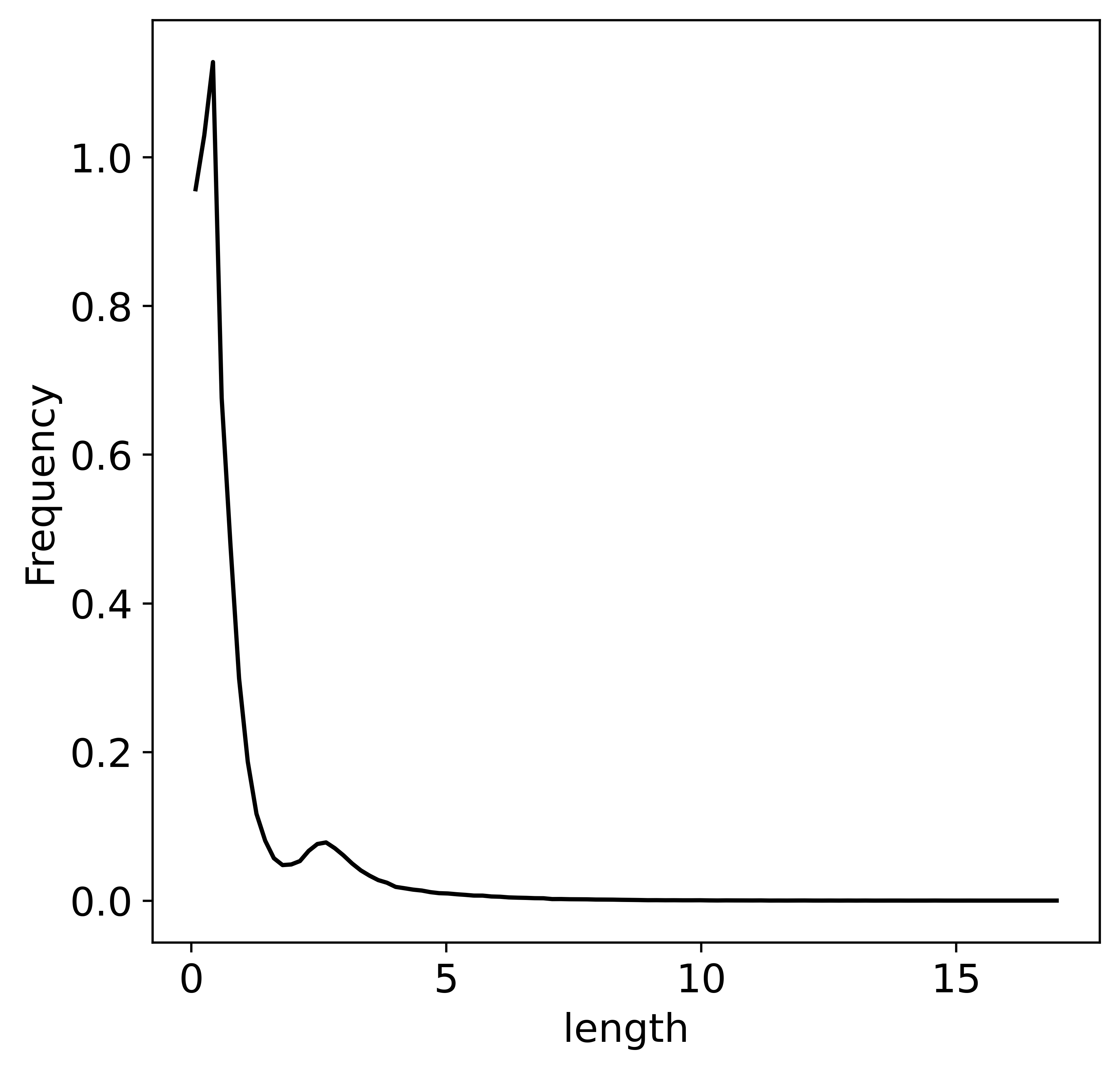}
        \caption{}
    \end{subfigure}
    \hfill
    \begin{subfigure}[b]{0.32\textwidth}
        \centering
        \includegraphics[width=\linewidth]{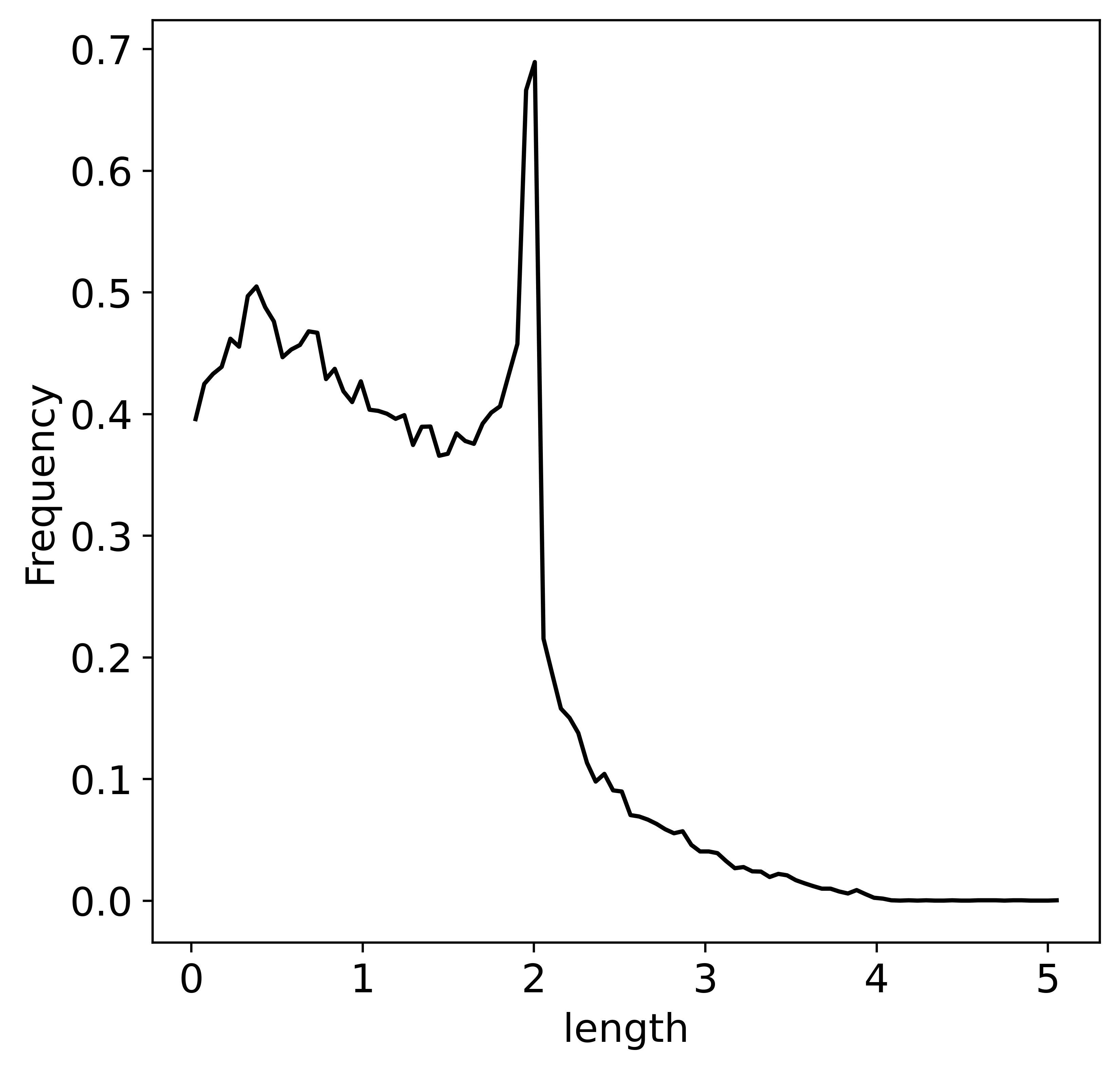}
        \caption{}
    \end{subfigure}
    \hfill
    \begin{subfigure}[b]{0.32\textwidth}
        \centering
        \includegraphics[width=\linewidth]{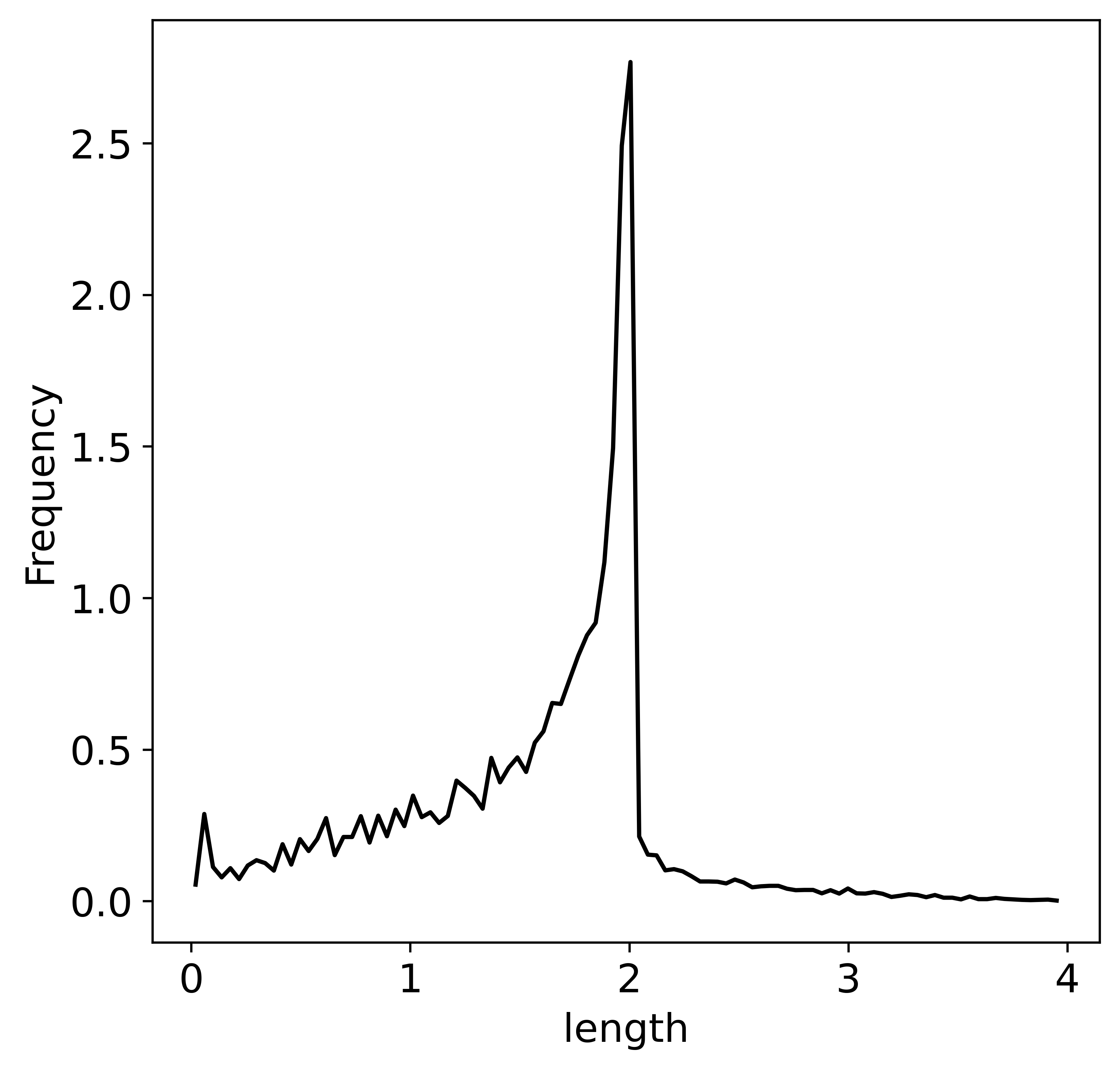}
        \caption{}
    \end{subfigure}
    \caption{Path length distribution for a process with memory, elastic boundaries, exponential step length distribution, $10^7$ collisions, 100 bins. Constant drift $C=\pi/8$. (a) $l/R=0.3$, (b) $l/R=1$, and (c) $l/R=5$.}
    \label{NEMO_prob}
\end{figure*}

The limit value of the mean path length as $l/R\rightarrow\infty$ is the Cauchy prediction, $\pi/2$, for all values of constant $C$. This occurs because the movement is essentially a billiard, and the influence of the memory process vanishes. The variance decreases because the probability distribution narrows around $l=2R$, as can be seen in Fig. (\ref{NEMO_prob}). In this type of memory process, the distribution has two peaks, one of which decreases while the other increases as $l/R$ increases. This demonstrates that memory processes introduce different movement patterns that deviate from the expected behavior under Markovian assumptions.

The memory process we chose is inspired by certain types of movement observed in bacteria, for example, the chiral movement seen in bacteria swimming close to a surface, as well as the behavior of some L-shaped active colloids. We selected this specific type of process primarily for its simplicity, which allows us to explore the fundamental aspects of memory in particle dynamics within a confined domain. However, we would like to clarify that this is not intended as a general result for all types of memory processes.

\begin{figure*}[!]
    \centering
    \begin{subfigure}[b]{0.33\textwidth}
        \centering
        \includegraphics[width=\linewidth]{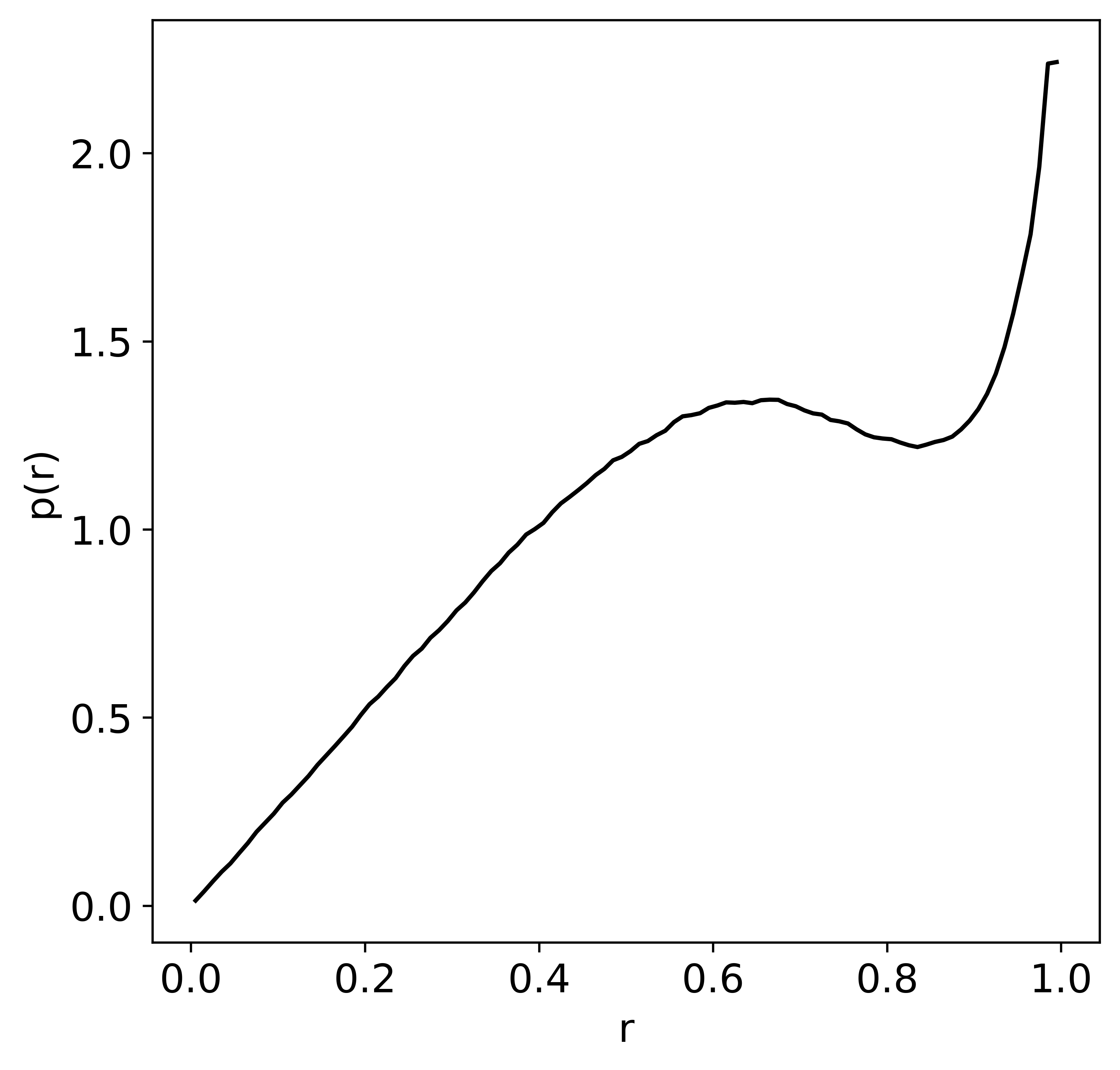}
        \caption{}
    \end{subfigure}
    \hfill
    \begin{subfigure}[b]{0.32\textwidth}
        \centering
        \includegraphics[width=\linewidth]{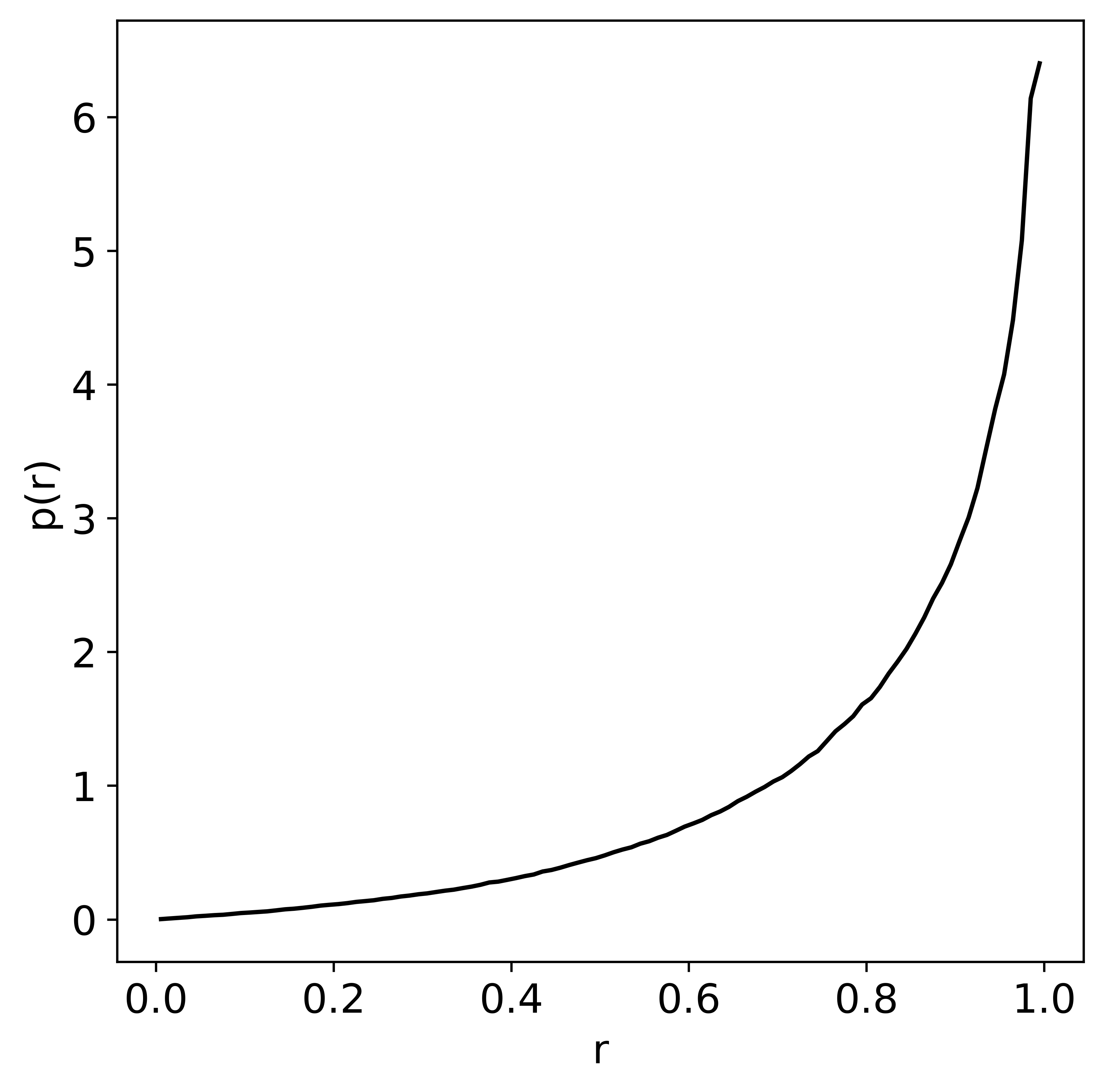}
        \caption{}
    \end{subfigure}
    \hfill
    \begin{subfigure}[b]{0.32\textwidth}
        \centering
        \includegraphics[width=\linewidth]{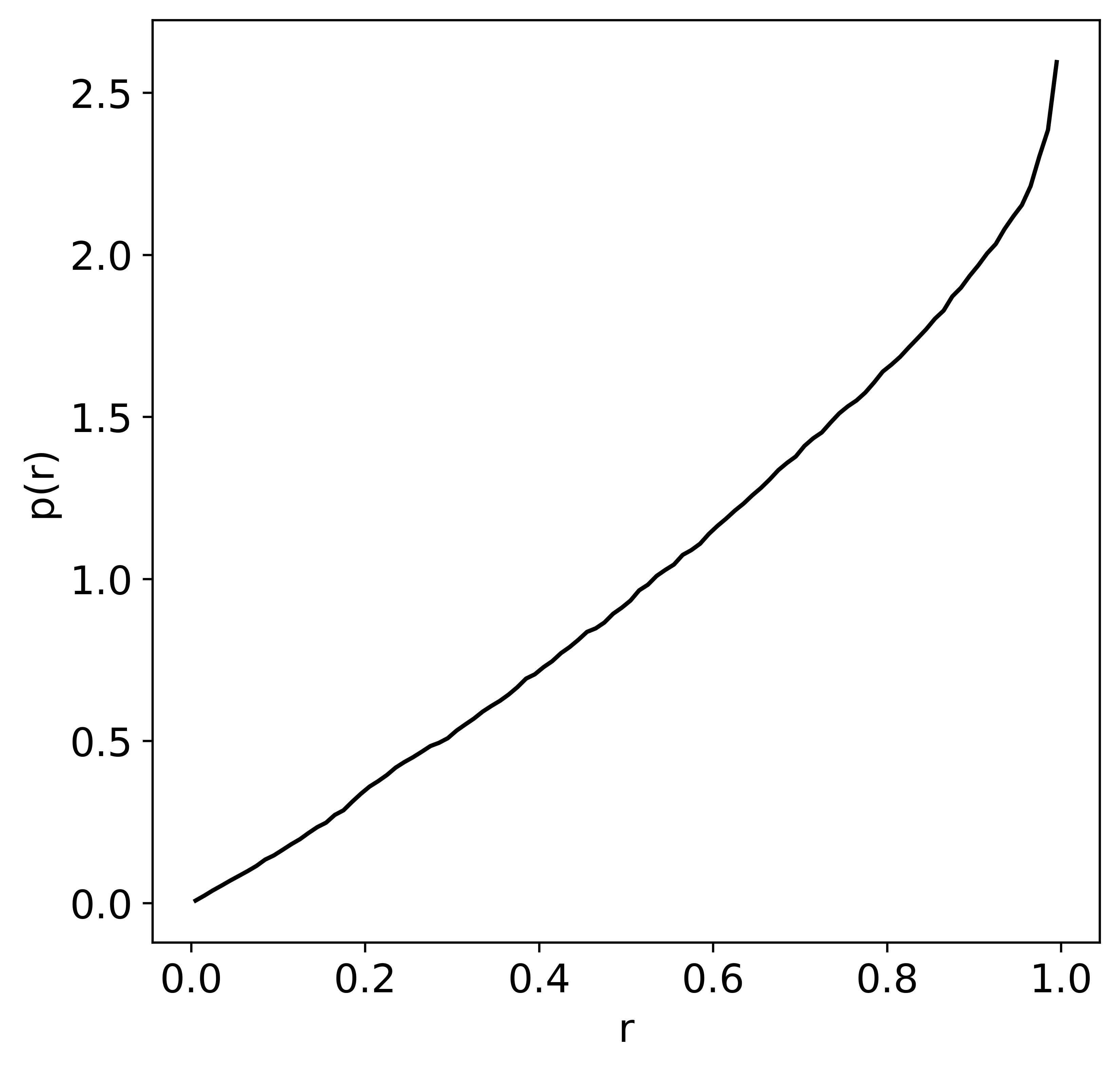}
        \caption{}
    \end{subfigure}
    \caption{Radial distribution for a process with memory, elastic boundaries, exponential step length distribution, $10^7$ collisions, 100 bins. Constant drift $C=\pi/8$. (a) $l/R=0.1$, (b) $l/R=0.5$, and (c) $l/R=1$.}
    \label{NEMO_radial}
\end{figure*}

Regarding the radial distribution, it presents a non-trivial form, as shown in Fig. (\ref{NEMO_radial}). This demonstrates how the reorientation mechanism of an active particle can produce distinctive distributions, thereby providing information about its underlying laws.

Understanding process memory is not only theoretically intriguing but also practically significant. For instance, Frangipane et al. demonstrated invariance properties of bacterial random walks in complex structures, suggesting that real-world bacterial motion often exhibits memory effects that need to be accounted for in predictive models \cite{Frangipane2019}. Additionally, the impact of memory on narrow escape times and the sorting of active particles was explored by Paoluzzi et al., emphasizing the practical applications in microfluidic device design and biological systems \cite{Paoluzzi2020}.

By integrating these insights, it becomes clear that memory effects play a fundamental role in the dynamics of confined run-and-tumble particles. This study’s exploration of simple memory processes provides a foundational understanding that can be expanded with more complex, real-world-inspired mechanisms in future research. Such efforts will enhance our ability to predict and control particle behavior in various scientific and technological applications.

\section{Conclusions}
\label{conclusions}

In this paper, we explored the dynamics of run-and-tumble motion within confined two-dimensional domains, specifically examining how various distributions and boundary conditions affect the validity of the Mean Path Length Theorem (MPLT) and the statistics of the dynamics. Through a combination of theoretical analysis and numerical simulations, we demonstrated that the MPLT holds true across a wide range of conditions, provided that the angular distribution of particle tumbling remains uniformly random. This underscores the significance of understanding motion within bounded domains across physics, ecology, and biology, and highlights the foundational role of the MPLT in predicting and controlling particle or agent behavior in diverse environments.

We used a simulation of a 2D circular domain to model particle movement, adhering to real-world experimental conditions. By employing a constant speed for particles and exploring various step length distributions, we observed the transition of movement from diffusive to quasi-ballistic. We then presented an analysis of the MPLT's applicability under diverse conditions, revealing that the theorem holds true when angular distributions are uniformly random. In fact, when particles exhibit a strong directional bias due to fixed or Gaussian-distributed angles, they tend to collide with the same region of the boundary, introducing anisotropy and thereby violating the MPLT's foundational assumptions.

We examined path length distributions and their variance under different conditions. Exploring path length distributions between collisions sheds light on the transition from diffusive to quasi-ballistic movements. This part of the study illustrates how varying the parameter $l/R$ influences the distribution, emphasizing the intricate relationship between movement type and path length distribution. The results show a transition from diffusive to quasi-ballistic regimes as $l/R$ increases. For $l/R<1$, the distribution is characterized by an exponential decay with a power-law deviation for small path lengths, transitioning to a combination of diffusive and ballistic behaviors as $l/R$ increases. Understanding these distributions helps in predicting escape times, optimizing sorting mechanisms, and designing environments that mimic natural habitats for microorganisms.

The analysis of radial distributions under different boundary conditions reveals a uniform density across all cases of $l/R$  when conditions are elastic, demonstrating a consistent linear radial distribution. The contrast with random angle boundary conditions, where the radial distribution deviates, highlights the significant impact of boundary interactions on particle distribution. This is relevant to other physically important problems of motion in bounded systems, such as the problem of narrow escape \cite{Holcman2016}, since increased spatial density near the edge raises the probability of hitting the target. Random reflections can also be important when considering reversible sticking to the boundary (see, e.g., \cite{Grebenkov2017, Grebenkov2022, Grebenkov2023}), because if a particle sticks to the boundary, it is natural to consider the outgoing angle of release as random and uncorrelated with the incoming direction. We show in this work that studying the statistics of movement in confined settings can help explain the mobility mechanisms of some active agents, like microbes. The way in which the active particle interacts with the boundary and the reorientation mechanism produces distinctive patterns of movement and distributions, giving us information about the underlying laws governing its movement.

The introduction of process memory into the study of run-and-tumble particles reveals critical deviations from traditional theoretical predictions, emphasizing the need to consider historical dependencies in particle motion. This aligns with and expands upon findings in the broader literature, underscoring the complex interplay between memory, boundary interactions, and particle dynamics in confined settings.

This work not only affirms the robustness of the MPLT under a broad spectrum of conditions but also highlights the importance of angular distribution and boundary interactions in governing the dynamics of confined motion. The implications of our research extend beyond theoretical physics, offering insights into biological processes such as bacterial chemotaxis and the design of microfluidic devices for sorting and analyzing microscopic particles.

\begin{acknowledgments}
R.A. acknowledges an affiliation with INDAM.
\end{acknowledgments}

\section*{Data Availability Statement}
The data that support the findings of this study are available within the article.

\section*{Author contribution statement}
D.J.Z.: Visualization, Validation, Investigation, Formal analysis, Data curation, Writing - original draft. R.A.: Writing - review \&
editing, Supervision, Resources, Methodology, Funding acquisition, Conceptualization.

\section*{Declaration of competing interest}
The authors have no conflicts to disclose.

\appendix

\section{Alternative calculation of distributions.}

We present here an alternative procedure for calculating the radial distribution based on \cite{Evans2001}. Consider the case of random angle boundary conditions. The distribution can be calculated as follows: consider a portion of the trajectory inside a circle of radius $R$, as shown in Fig. (\ref{vectors}). The particle is situated at the point $x$, at distance $r$ from the center, and moves in the direction $\overrightarrow{v}$, where $|\overrightarrow{v}|=1$. In this situation, the particle will follow the trajectory $\tau$, and will collide with the boundary at the point $y$. $v_y$ is the normal vector at the point $y$, as shown in Fig. (\ref{vectors}).

Following \cite{Evans2001}, the probability of the particle of being in a given point with a certain direction for the case of random boundary conditions is:

\begin{equation}
	\Pi(x,\theta)=\frac{\alpha}{|\overrightarrow{v}\cdot\overrightarrow{v_y}|},
\end{equation}

\nd where $\alpha$ is a normalization constant. The scalar product of $\overrightarrow{v}$ and $\overrightarrow{v_y}$ is

\begin{equation}
	\overrightarrow{v}\cdot\overrightarrow{v_y}=|\overrightarrow{v}|\times|\overrightarrow{v_y}|\times\cos\phi,
\end{equation}

\nd and using the sine theorem,

\begin{equation}
	\frac{r}{\sin\phi}=\frac{R}{\sin\theta},
\end{equation}

\nd we obtain

\begin{equation}
	|\cos\phi|=\sqrt{1-\sin^2\phi}=\sqrt{1-\frac{r^2}{R^2}\sin^2\theta}.
\end{equation}

Therefore,

\begin{equation}
	\Pi(x,\theta)=\frac{\alpha}{\sqrt{1-\frac{r^2}{R^2}\sin^2\theta}}.
\end{equation}

\begin{figure}[!]
    \centering
    \includegraphics[width=.7\linewidth]{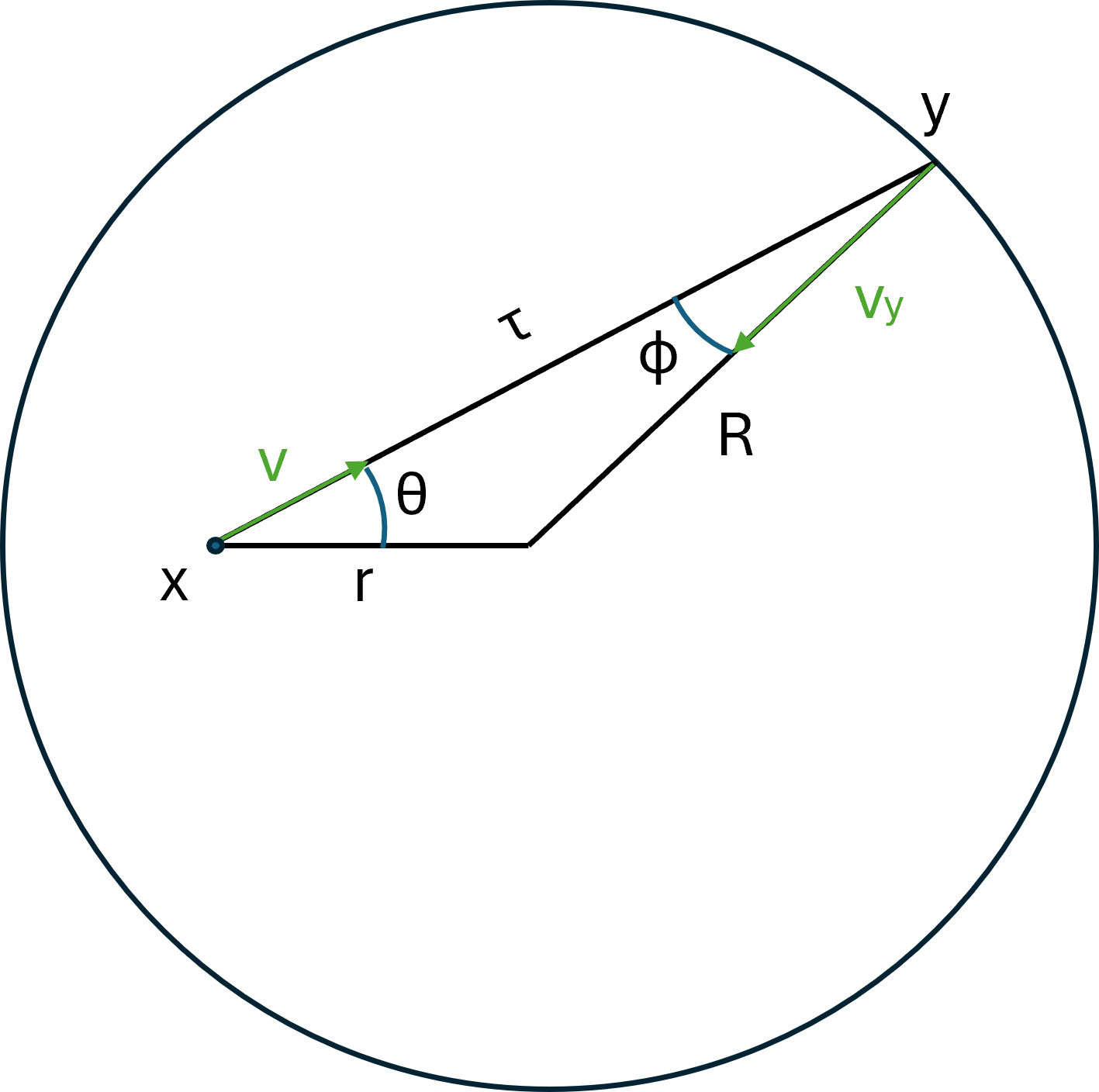}
    \caption{Geometry of a single collision with the boundary.}
    \label{vectors}
\end{figure}

The probability of being at the point $x$ is

\begin{equation}
\begin{split}
	\rho(x)&=\int_0^{2\pi}\frac{\alpha d\theta}{\sqrt{1-\frac{r^2}{R^2}\sin^2\theta}}=4\alpha\int_0^{\pi/2}\frac{\alpha d\theta}{\sqrt{1-\frac{r^2}{R^2}\sin^2\theta}}\\
	&=4\alpha K\left(\frac{r}{R}\right),
\end{split}
\end{equation}

\nd where $K(x)$ is the complete elliptic integral of the first kind, defined as

\begin{equation}
K(x) = \int_0^{\pi/2}\frac{ds}{\sqrt{1-x^2\sin^2s}}.
\end{equation}

Finally, the radial probability is

\begin{equation}
	p(r)=2\pi r\times4\alpha K\left(\frac{r}{R}\right).
\end{equation}

After normalization, we find that, $\alpha=1/8\pi R^2$, so

\begin{equation}
	p(r)=\frac{r}{R^2}K\left(\frac{r}{R}\right),
\end{equation}

\nd the same results as in eq. (\ref{randomeq}). 

A similar calculation can be performed for the elastic case. For elastic collisions $\Pi(x,\theta)=\alpha$, since the invariant measure is uniform, and the computation is simple.

\begin{equation}
	\rho(x)=\int_0^{2\pi}\alpha d\theta=2\pi\alpha.
\end{equation}

The radial probability is

\begin{equation}
	p(r)=2\pi r\times 2\pi\alpha=4\pi\alpha r,
\end{equation}

\nd which after normalization becomes

\begin{equation}
	p(r)=\frac{2r}{R^2},
\end{equation}

\nd the same result as in eq. (\ref{radialeq}).

\nocite{*}
\bibliography{run-and-tumble}

\end{document}